\documentclass[pre,twocolumn,showpacs,preprintnumbers,superscriptaddress,floatfix,nofootinbib]{revtex4-1}
\usepackage{subfigure}
\usepackage{amsfonts}
\usepackage{graphicx,,booktabs}
\usepackage{amssymb}
\usepackage{amsmath,txfonts,ulem}
\usepackage{graphicx}
\usepackage{epstopdf}
\usepackage{epsfig}
\usepackage{dcolumn}
\usepackage{color}
\usepackage{bm}
\usepackage{multirow}
\usepackage{ulem}
\usepackage[colorlinks,
citecolor=blue,
anchorcolor=green,
menucolor=orange,
linkcolor=red,
filecolor=red,
runcolor=pink,
urlcolor=blue,
frenchlinks=red]{hyperref}

\allowdisplaybreaks[4]
\begin{document}
	
	\title{Synchronous control study of   Chua circuit system via capacitive closed-loop coupling}

	\author{Jiyuan Zhang}
	\email{jyuanzhang@yeah.net}
	\affiliation{Department of physics, Lanzhou University of Technology,
		Lanzhou 730050, China}

	\author{Xiao-Yun Wang}
	\email{xywang@lut.edu.cn}
	\affiliation{Department of physics, Lanzhou University of Technology,
		Lanzhou 730050, China}
	\affiliation{Lanzhou Center for Theoretical Physics, Key Laboratory of Theoretical Physics of Gansu Province, Lanzhou University, Lanzhou, Gansu 730000, China}

	\begin{abstract}
		Synchronous control of nonlinear circuits is of great importance in many fields. In this paper, a capacitor is used for closed-loop coupling of three dual-vortex attractor Chua circuits with the same circuit parameters and different initial conditions, and the corresponding synchronization processes and synchronization effects are investigated. It is found that, for the capacitor, the closed-loop coupling can completely synchronize the three Chua circuits. And there exists a critical coupling strength $\delta_{critical}^C$ that can be calculated, and when the coupling strength is between the critical coupling strength $\delta_{critical}^C$ and the upper coupling strength $\delta_{max}^C$, the three circuits can be completely synchronized very quickly. Moreover, the results show that the unidirectional coupling cannot make the circuit system completely synchronized. Finally, we also verify the correctness of the calculation by circuit simulation experiments.

	\end{abstract}
	
	
	\maketitle
	
	\section{Introduction}
	In traditional circuit theory, circuits are usually classified into two main categories: linear and nonlinear circuits. In fact there is no strictly linear circuit, it only appears to be approximately linear around a specific operating point. Therefore, the study of nonlinear circuits is particularly important. A nonlinear circuit is a circuit containing nonlinear components. With the development of electronics, many highly nonlinear electronic devices have emerged, such as amnesic resistors \cite{ref1,ref2}, nonlinear diodes \cite{ref3}, and so on. This makes the nonlinear circuits composed by these devices widely used in electronic circuits, and also makes a large number of nonlinear phenomena appear in electronic circuits, such as self-excited oscillations, probable periodic motion \cite{ref4} and chaos \cite{ref5,ref6}. So are these nonlinear phenomena beneficial or harmful in practical use? Does it have any exploitable value? Take chaotic phenomena as an example, the unpredictability and initial value sensitivity of chaotic systems make it difficult to fully control chaotic phenomena. However, the control of chaotic phenomena can be applied to many scenarios, such as secure communication \cite{ref7},  information processing\cite{ref8},  and even biological neuronal analysis\cite{ref9,ref10,ref11} in which would like to control chaotic phenomena. L. M. Pecora and T. L. Carroll proposed a method for analyzing identical chaotic systems using principal stability functions \cite{ref12,ref13,ref14}. Chua circuits a simple nonlinear electronic circuit design which can exhibit standard chaos theory behavior\cite{ref15}. The Chua circuit can be accurately modeled mathematically and is considered as a "paradigm of chaotic systems". Therefore, it may be a good choice to start from the synchronous control of Chua circuits to further study chaotic control.

	In recent years, many approaches to the control of Chua circuits have been proposed. As the founders of Chua circuits, L. O. Chua et al. first investigated the synchronization between Chua circuits experimentally\cite{ref16}. Chaotic control by time lag feedback controller\cite{ref17} and adaptive feedback controller \cite{ref18}is also the mainstream control method nowadays. By improving the basic Chua circuit and studying the pulse synchronization and phase synchronization in it\cite{ref19,ref20}, derived some methods to make the Chua circuit self-regulating. In practice,  most nonlinear circuit models are in the form of systems, and a Chua circuit can only be considered as a unit in a system. Therefore, by activating coupling channels between different circuit modules, they can be regulated to each other. This approach would be more relevant to the real scenario.

	Based on this idea, Ref. \cite{ref21} successfully achieved simultaneous regulation of two Chua by activating a capacitive coupling channel between two Chua circuits and exploiting the feature of capacitors to store energy. Similarly, Ref. \cite{ref22} investigated the synchronization between two chua circuits coupled through an inductor; Ref. \cite{ref23} investigated the synchronization between two chua circuits coupled through a thermistor. It was found that these nonlinear components have a good modulation effect between both chua circuits, and the concept of coupling strength was given to characterize the magnitude of the coupling ability between two circuits by different components \cite{ref21,ref22,ref23}. More complex, the same control can be achieved by coupling through amnesic resistors or other combinations of nonlinear elements \cite{ref24,ref25,ref26}. With these research foundations, Ref. \cite{ref27} studied the synchronization of a circuit system consisting of fifty Chua circuits with different nonlinear elements coupled unidirectionally. The results show that the coupling strengths corresponding to different elements vary considerably in value and that the coupling strength directly determines the synchronization effect between circuits. For systems where the number of circuits to be coupled is greater than or equal to three, the control method is divided into open-loop control (no feedback control) or feedback control for implementation. The previously mentioned unidirectional coupling becomes a kind of feedback-free control. In fact, in contrast to feedback-free control, feedback control allows the system state to move away from chaotic motion and into an arbitrary desired immobility point. This makes it easier to synchronize the multi-circuit system completely. Therefore, our work builds on the former study and proposes and computationally proves a closed-loop coupling of multiple nonlinear circuits using the basic idea of feedback control.

	The main body of this paper is divided into the following parts: First, in Sec. \ref{sec1}, we start with a basic Chua circuit, which is identically coupled to three chua circuits by capacitive  closed-loop. And the corresponding circuit equations as well as the dimensionless equations are given.The parameters chosen for our subsequent calculations are also given. Second, in Sec. \ref{sec2}, we fix the Chua circuit parameters and give the initial conditions for each of the three Chua circuits to be impassable. Based on this, we calculated the synchronization process and synchronization effect of the circuits when they are connected through the closed-loop capacitor, by adjusting the values of the coupling strength between the circuits. And the results are compared with the synchronization results of unidirectional coupling, and the advantages of closed-loop coupling are derived. In Sec. \ref{sec3}, we build a simulation circuit and observe the experimental results to further prove the feasibility of the theory and the numerical realism. Finally, a summary is given in Sec. \ref{sec4}.

	\section{Circuit model}
	\label{sec1}
	\subsection{Chua circuit and parameter selection}
	
	Chua circuits are the simplest class of nonlinear circuits with the circuit equation:
	\begin{equation}
		\begin{cases}
			C_1\frac{dV_{C_1}}{dt}=\frac{V_{C_2}-V_{C_1}}{R}-f(V_{C_1})\\
			C_2\frac{dV_{C_2}}{dt}=\frac{V_{C_1}-V_{C_2}}{R}+i_L\\
			L\frac{di_L}{dt}=-V_{C_2}
		\end{cases}
	\end{equation}
	where
	\begin{equation}
		f(V_C)=G_aV_C+0.5(G_a-G_b)(|V_C+E|-|V_C-E|)
	\end{equation}
	is the characteristic function of the Chua nonlinear diode. where $G_a$, $G_b$, and $E$ are constants, all of which are characteristic properties of the element itself. By the proportional transformation of the Chua circuit \cite{ref28}
	
	\begin{equation}
		\begin{aligned}
			&x=\frac{V_{C_1}}{E},y=\frac{V_{C_2}}{E},z=\frac{i_LR}{E},\tau=\frac{t}{RC_2}\\
			&m_0=RG_a,m_1=RG_b,\alpha=\frac{C_2}{C_1},\beta=\frac{C_2R^2}{L}
		\end{aligned}
		\label{equ:3}
	\end{equation}
	
	The dimensionless kinetic equations of a single Chua circuit can be obtained \cite{ref15}
	\begin{equation}
		\begin{cases}
			\dot{x}=\alpha[y-x-f(x)]\\
			\dot{y}=x-y-z\\
			\dot{z}=-\beta y
		\end{cases}
	\end{equation}
	
	where
	\begin{equation}
		f(x)=m_{1} x+0.5\left(m_{0}-m_{1}\right)(|x+1|-|x-1|)
	\end{equation}

	The parameters of the Chua circuit are chosen as\cite{ref16}
	\begin{equation}
		\begin{aligned}
			&C_{1}  =10 \mathrm{nF}, C_{2}=100 \mathrm{nF}, E=1 V, L=18.75 \mathrm{mH} \\
			&G =\frac{1}{R}=0.599 \mathrm{mS}, G_{a}=-0.76 \mathrm{mS}, G_{b}=-0.41 \mathrm{mS}
		\end{aligned}
		\label{equ:6}
	\end{equation}
	The corresponding dimensionless parameters are
	\begin{equation}
		\alpha=10 , \beta=14.87 , a=-1.27, b=-0.68
		\label{equ:7}
	\end{equation}
	
	A set of parameters such as these triggers the chaos of the Chua circuit, and our subsequent calculations, simulations, and discussions are based on these parameters.
	
	In the subsequent study, We will discuss the closed-loop coupling of the "$x$" outputs of the three Chua circuits through capacitors and discuss the synchronization process between the circuits.
	
	\subsection{Capacitive Closed-loop Coupling}
	\label{sec:2}
\begin{figure*}[htbp]
		\centering
		\includegraphics[width=1.0\linewidth]{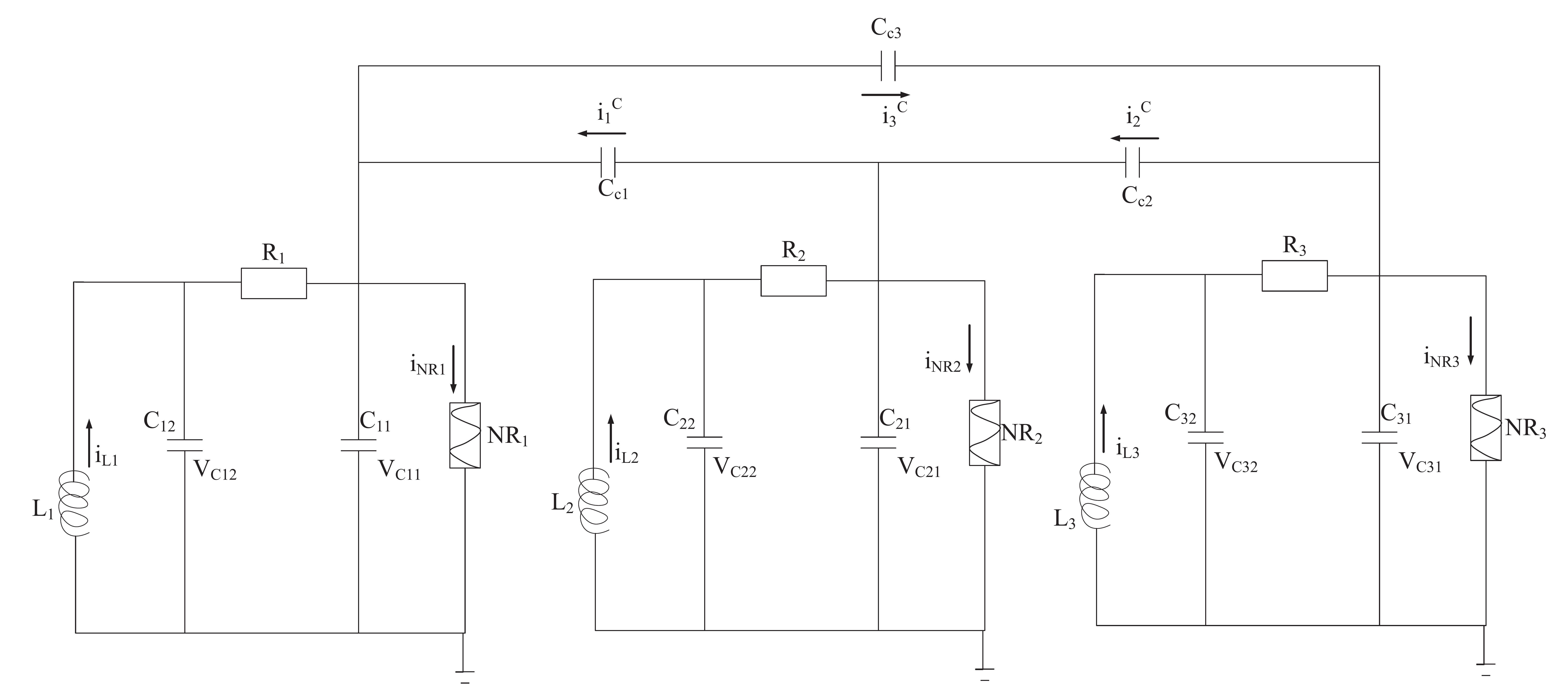}
		\caption{Capacitive closed-loop coupling three Chua circuit schematic}
		\label{fig:3}
	\end{figure*}

	First, the three Chua circuits are closed-loop coupled through capacitor $C_c$ (Fig. \ref{fig:3}). The circuit equation is:

	\begin{equation}
		\begin{cases}
			C_{11}\frac{dV_{C_{11}}}{dt}=\frac{V_{C_{12}}-V_{C_{11}}}{R_1}-f(V_{C_{11}})+i_1^C-i_3^C\\
			C_{12}\frac{dV_{C_{12}}}{dt}=\frac{V_{C_{12}}-V_{C_{12}}}{R_1}+i_{L_1}\\
			L_1\frac{di_{L_1}}{dt}=-V_{C_{12}}\\
			
			C_{21}\frac{dV_{C_{21}}}{dt}=\frac{V_{C_{22}}-V_{C_{21}}}{R_2}-f(V_{C_{21}})+i_2^C-i_1^C\\
			C_{22}\frac{dV_{C_{22}}}{dt}=\frac{V_{C_{21}}-V_{C_{22}}}{R_2}+i_{L_2}\\
			L_2\frac{di_{L_2}}{dt}=-V_{C_{22}}\\
			
			C_{31}\frac{dV_{C_{31}}}{dt}=\frac{V_{C_{32}}-V_{C_{31}}}{R_3}-f(V_{C_{31}})+i_3^C-i_2^C\\
			C_{32}\frac{dV_{C_{32}}}{dt}=\frac{V_{C_{32}}-V_{C_{32}}}{R_3}+i_{L_3}\\
			L_3\frac{di_{L_3}}{dt}=-V_{C_{32}}
		\end{cases}
		\label{equ:13}
	\end{equation}

	The current through the coupling capacitor is $i_1^C $ , $i_2^C$ , $i_3^C$. Since the capacitor will continuously show charge and discharge in the circuit, it is also called feedback current. The capacitor will apply an feedback current on the Chua circuit to which it is connected. The specific form is:
	\begin{equation}
		\left\{\begin{array}{l}
			i_{1}^C=C_{c_1}\left(\dot{V}_{C_{21}}-\dot{V}_{C_{11}}\right)\\
			i_{2}^C=C_{c_2}\left(\dot{V}_{C_{31}}-\dot{V}_{C_{21}}\right)\\
			i_{3}^C=C_{c_3}\left(\dot{V}_{C_{11}}-\dot{V}_{C_{31}}\right)
		\end{array}\right.
	\end{equation}

	where $C_{c_k} , (k=1,2,3)$ is the capacitance of the coupling capacitor,. Similarly, the dimensionless equation for a system coupled through a closed-loop capacitor can be obtained by the dimensional transformation Eq. (\ref{equ:3}).
	\begin{equation}
		\begin{cases}
			\dot{x_1}=\alpha_1[y_1-x_1-f(x_1)]+\delta_1^C(\dot{x_2}-\dot{x_1})-\delta_3^C(\dot{x_1}-\dot{x_3})\\
			\dot{y_1}=x_1-y_1-z_1\\
			\dot{z_1}=-\beta_1 y_1\\
			
			\dot{x_2}=\alpha_2[y_2-x_2-f(x_2)]+\delta_2^C(\dot{x_3}-\dot{x_2})-\delta_1^C(\dot{x_2}-\dot{x_1})\\
			\dot{y_2}=x_2-y_2-z_2\\
			\dot{z_2}=-\beta_2 y_2\\
			
			\dot{x_3}=\alpha_3[y_3-x_3-f(x_3)]+\delta_3^C(\dot{x_1}-\dot{x_3})-\delta_2^C(\dot{x_3}-\dot{x_2})\\
			\dot{y_3}=x_3-y_3-z_3\\
			\dot{z_3}=-\beta_3 y_3
		\end{cases}
		\label{equ:15}
	\end{equation}
	where$\delta_1^C$, $\delta_2^C$ , $\delta_3^C$ is the coupling constant at capacitive coupling, defined as \cite{ref21}

	\begin{equation}
		\delta_k^C=\frac{C_{c_k}}{C_{k1}+2C_{c_k}}\quad k=1,2,3
		\label{equ:11}
	\end{equation}
	
	The value of the capacitive coupling strength is changed  by changing the value of the capacitance at the output of the Chua circuit and the value of the coupling capacitance. And the feedback current itself is also time-varying.Also, the strength of the coupling directly determines whether the circuits can be synchronized with each other.
	
	Next, we will discuss in detail the synchronization process between the three Chua circuits when they are closed-loop coupled through capacitors.
	
	\section{Numerical results}
	\label{sec2}
	\subsection{Synchronous Control Function Discussion}
	\label{sec:3}
\begin{figure*}[htbp]
		\centering
		\begin{tabular}{ccc}
			\includegraphics[width=60mm]{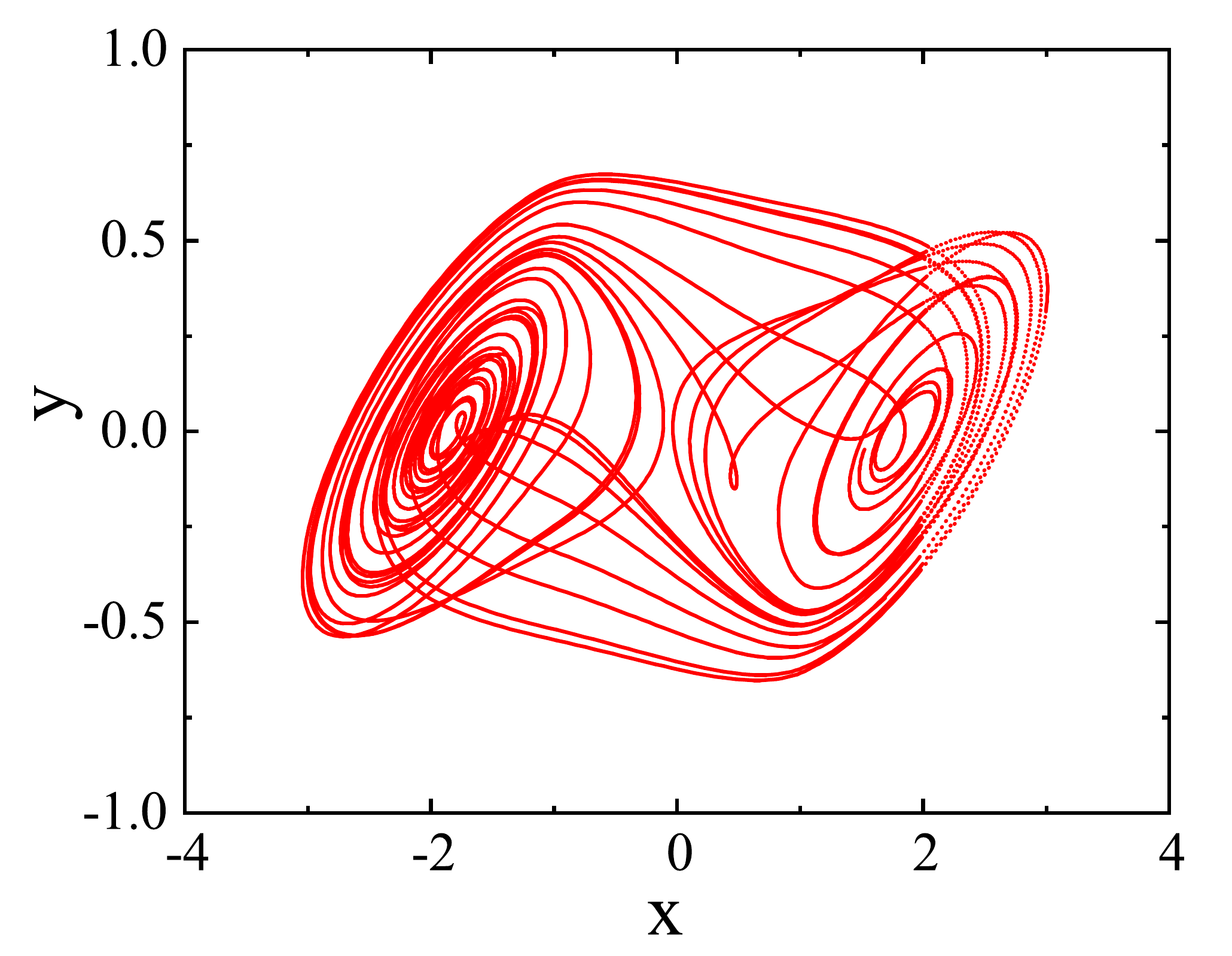}
			\includegraphics[width=60mm]{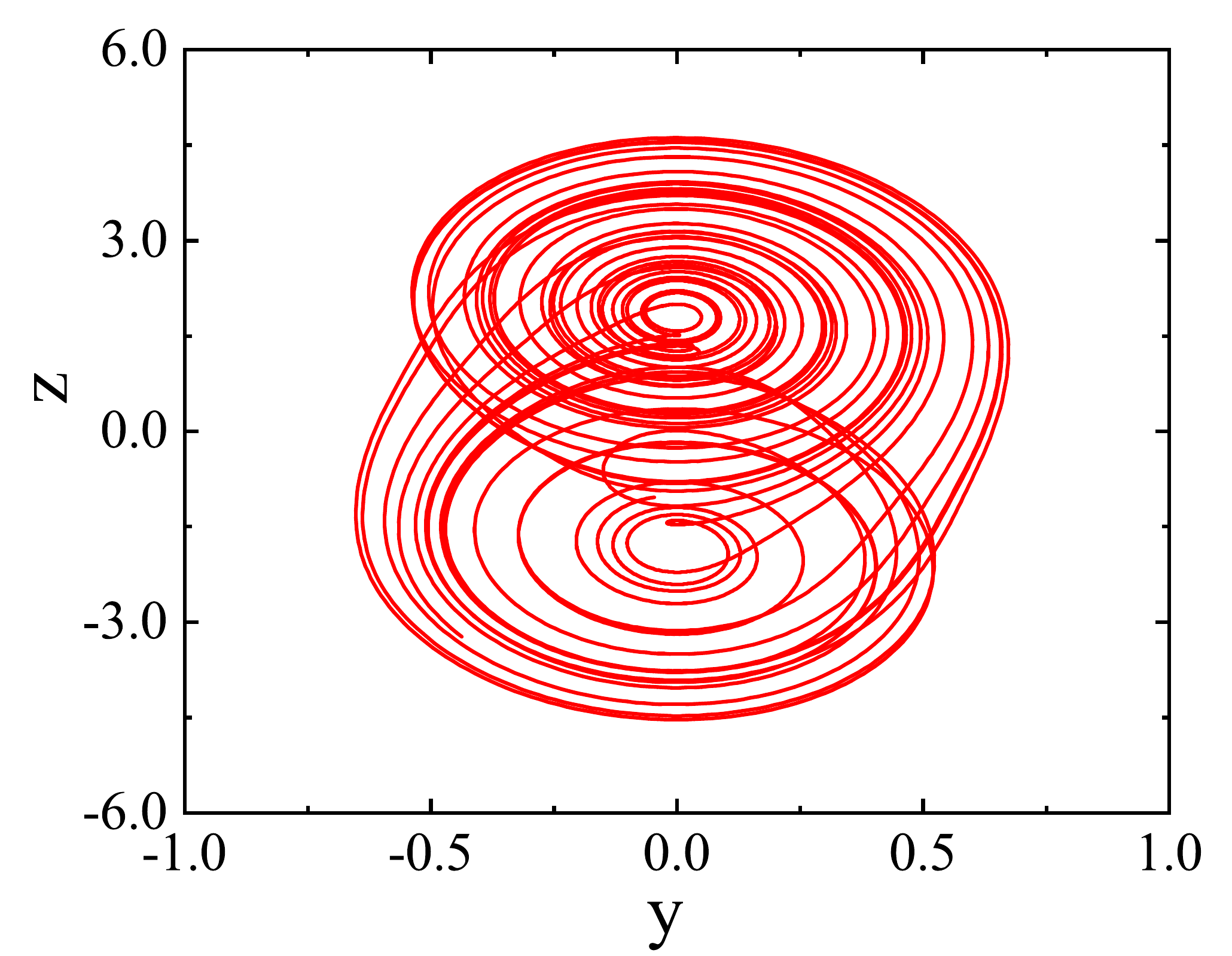}
			\includegraphics[width=60mm]{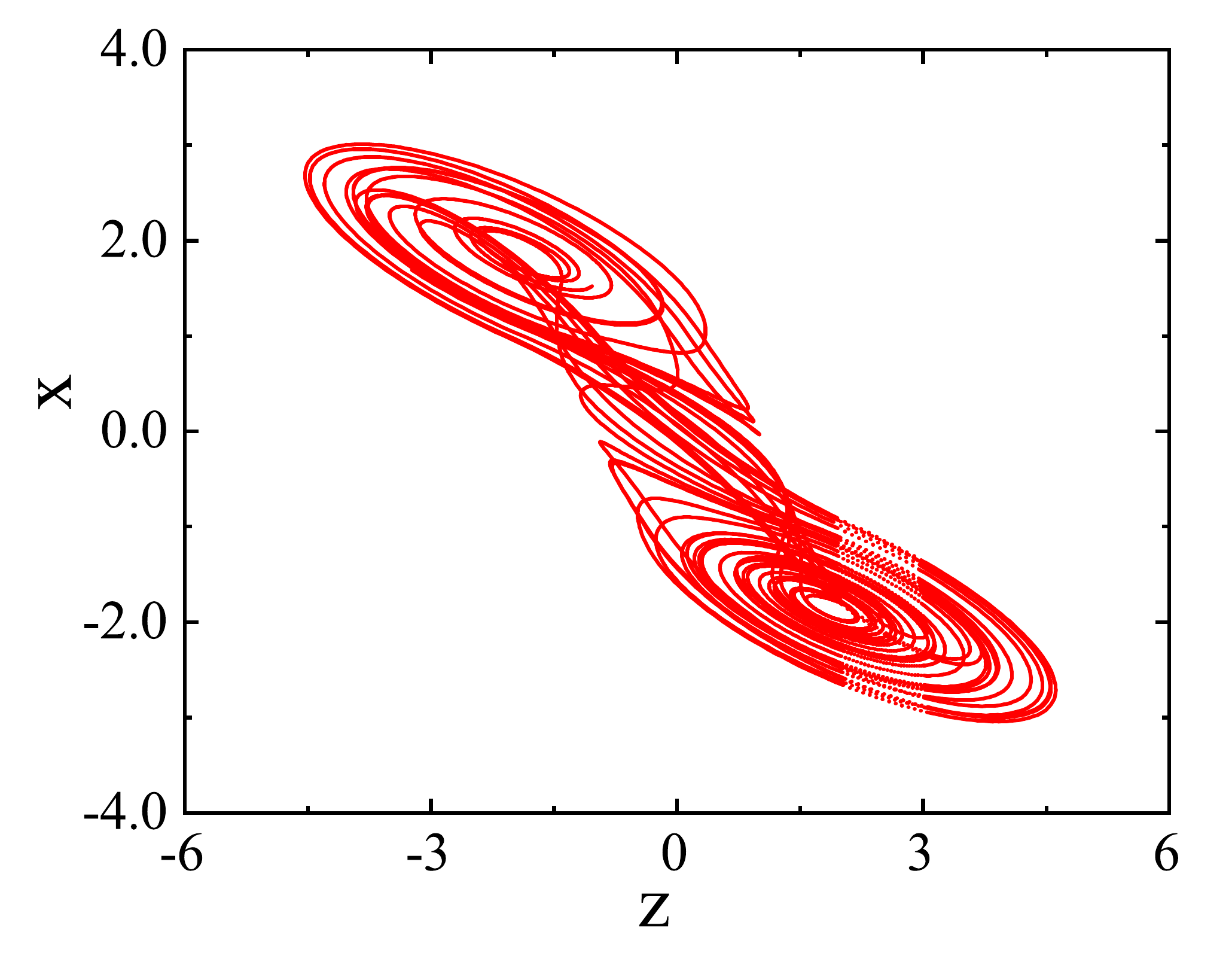}
		\end{tabular}
		\caption{  {\footnotesize  Dual vortex attractor of Chua circuit with parameter set to Eq. (\ref{equ:7}) .}}
		\label{fig:a5}
	\end{figure*}

	The output, oscillations and nonlinearities of the Chua circuit can be adjusted by setting different parameters with initial conditions for the set of dimensionless equations. In the subsequent discussion, we uniformly choose the parameters in Eq. (\ref{equ:7}) as the coefficients in the dimensionless equations. Such a set of parameters allows the Chua circuit to excite a double vortex attractor as shown in Fig. \ref{fig:a5} . Using the control variables, we set different initial conditions for the three Chua circuits discussed, as follows: $[x_1(0),y_1(0),z_1(0)]=[0.5,0.1,0.1]$ , $[x_2(0),y_2(0),z_2(0)]=[0.1,0.5,0.1]$ , $[x_3(0),y_3(0),z_3(0)]=[0.1,0.1,0.5]$

	First, we discuss whether the three Chua circuits can be synchronized when they are closed-loop coupled using different energy storage elements. We use the error function between the two circuits to characterize the degree of synchronization between them. The error function is defined as \cite{ref29,ref30}:
	\begin{equation}
		E_{ij}=\sqrt{(x_i-x_{j})^{2}+(y_i-y_{j})^{2}+(z_i-z_{j})^{2}} 
	\end{equation}
	where $i,j$ is the ordinal number of the two different Chua circuits and $i \neq j$. $E_{ij}$ denotes the synchronization error between the $i$-th circuit and the $j$-th circuit.We consider that the two Chua circuits are fully synchronized when $E_{ij} \rightarrow 0$. The three Chua circuits as a whole reach complete synchronization when every two circuits are synchronized with each other.
	
\begin{figure*}[ht]
		\centering
		\begin{tabular}{ccc}
			\includegraphics[width=60mm]{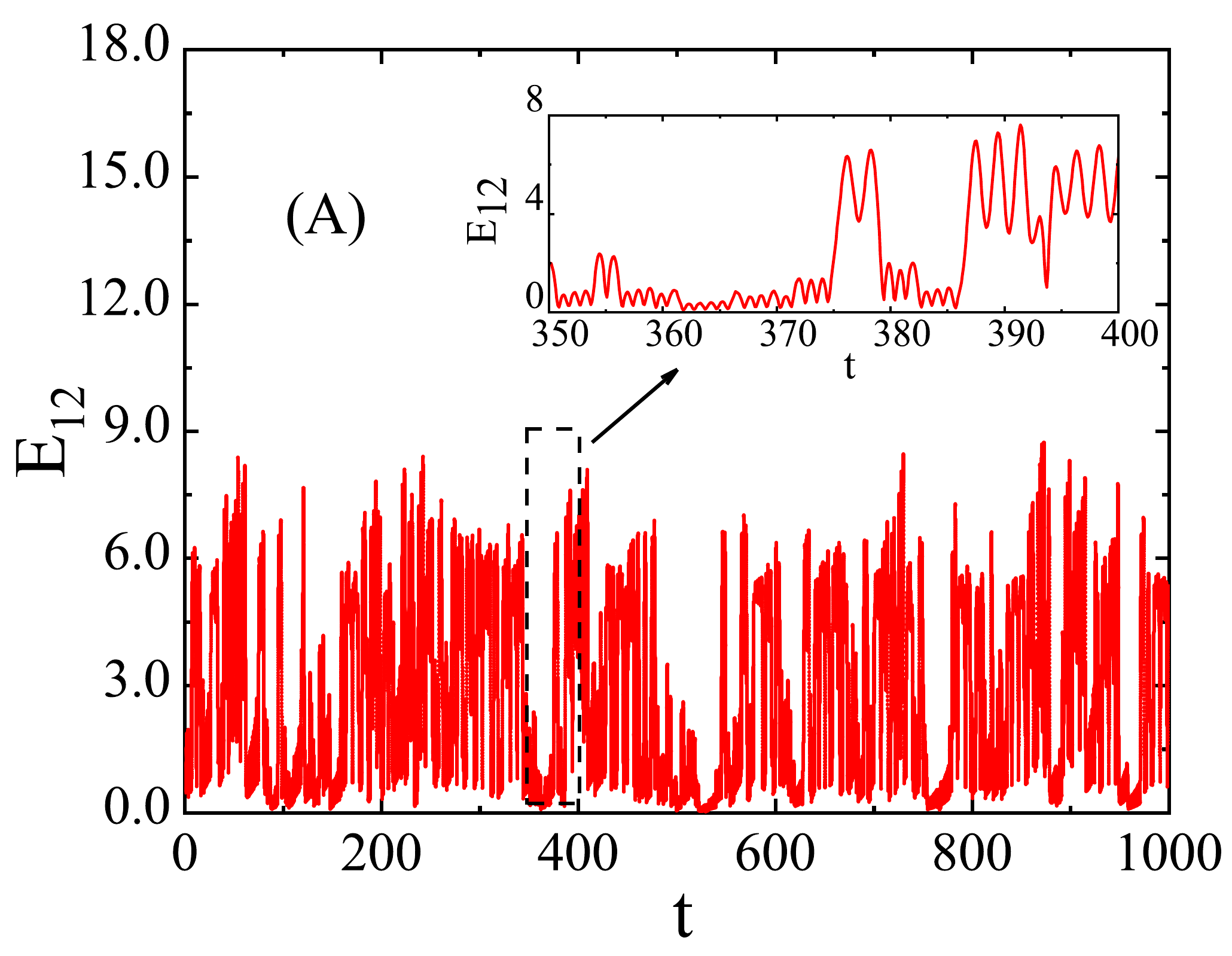}
			\includegraphics[width=60mm]{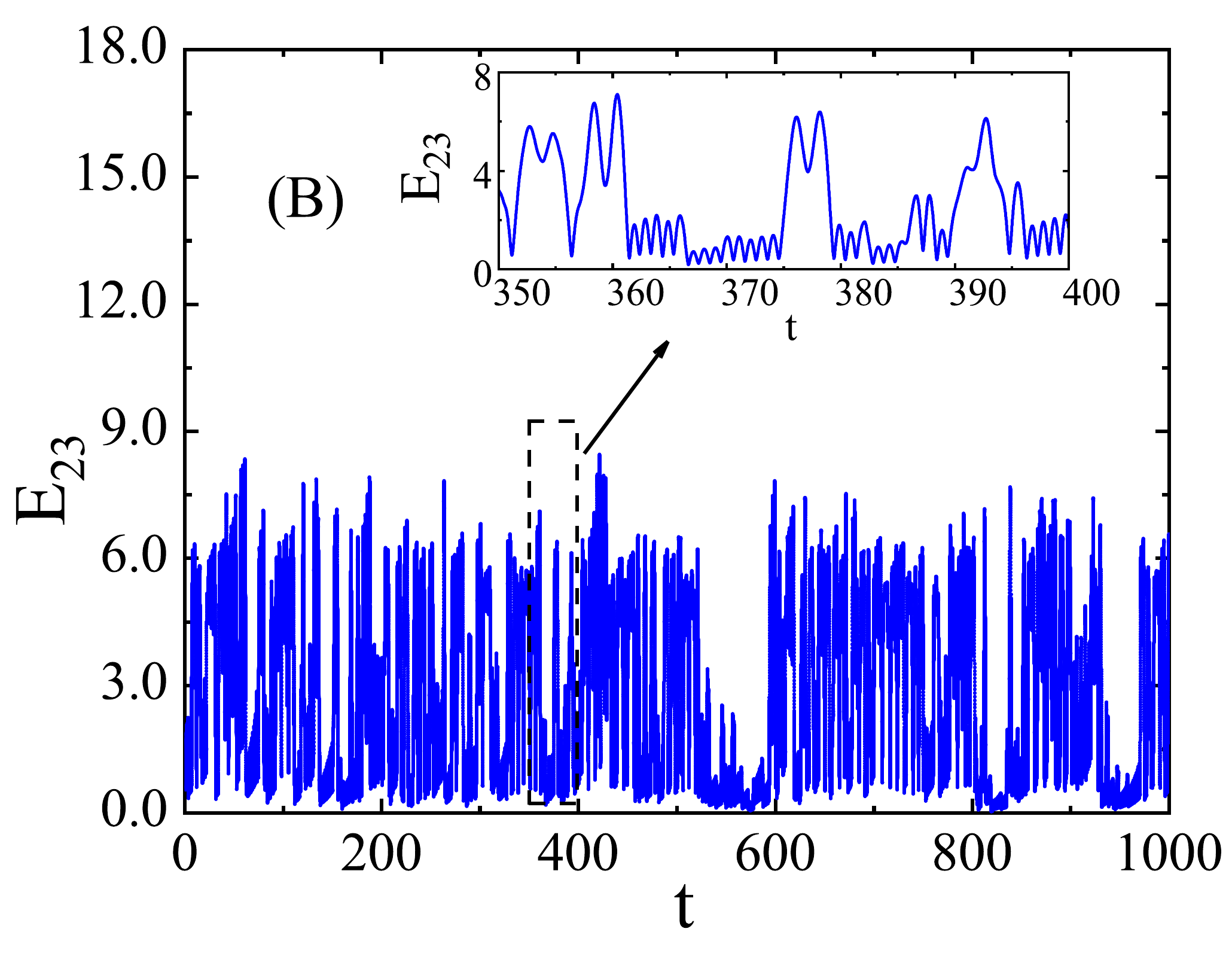}
			\includegraphics[width=60mm]{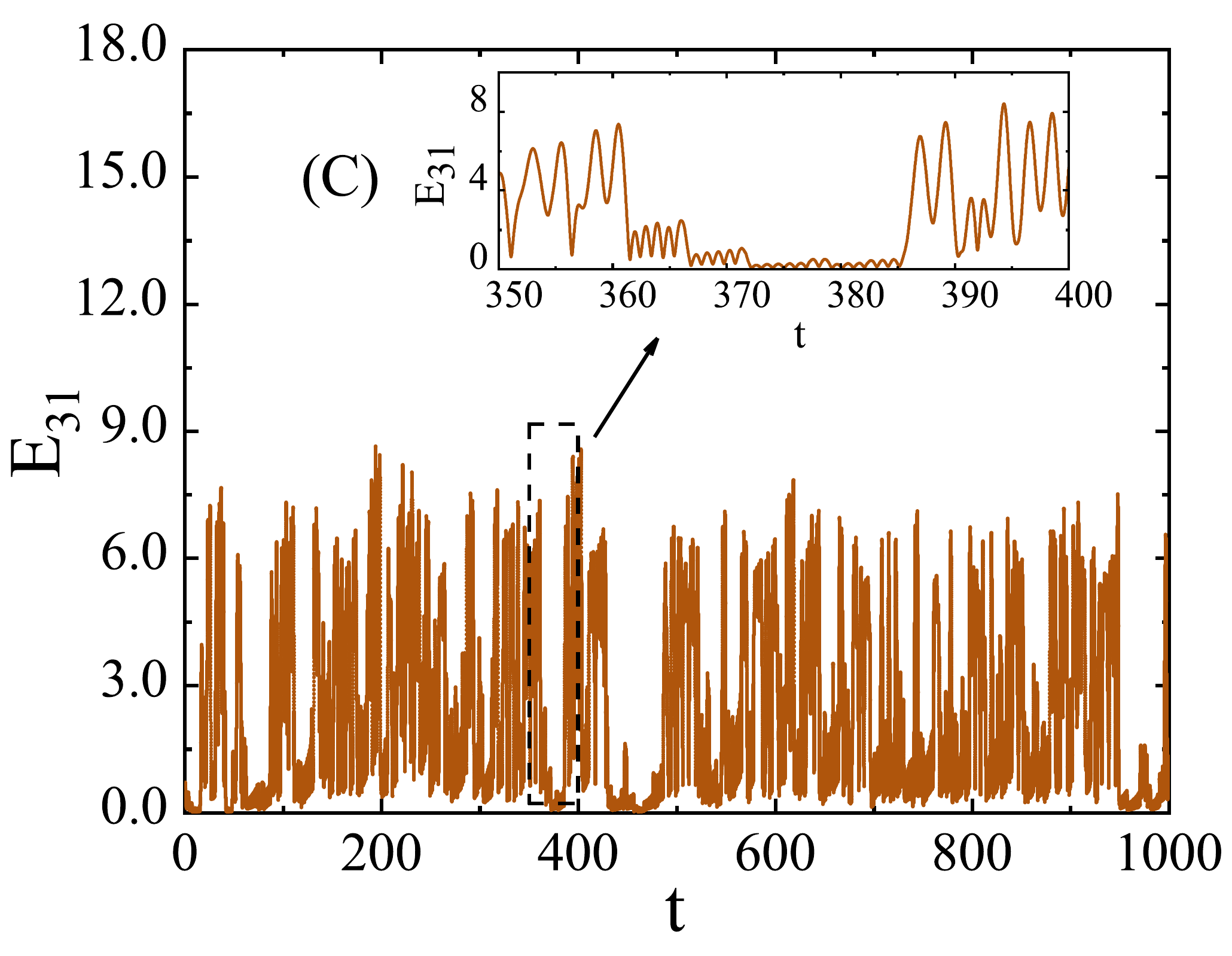}
		\end{tabular}
		\caption{The evolution of synchronization error between circuits coupled by closed-loop capacitance for coupling strength $\delta_1^C=\delta_2^C=\delta_3^C=0.1$  .}
		\label{fig:5}
	\end{figure*}

	For capacitive closed-loop coupling of the three Chua circuits. First we set the coupling strength $\delta_1^C=\delta_2^C=\delta_3^C=0.1$ . The corresponding capacitance value of the coupling capacitor $C_{c_1}=C_{c_2}=C_{c_3}=1.25 $ nF. The error evolution between the three circuits at this point is shown in Fig. \ref{fig:5}.

\begin{widetext}
		\begin{figure}[h]
			\centering
			\begin{tabular}{cccccc}
				\includegraphics[width=61mm]{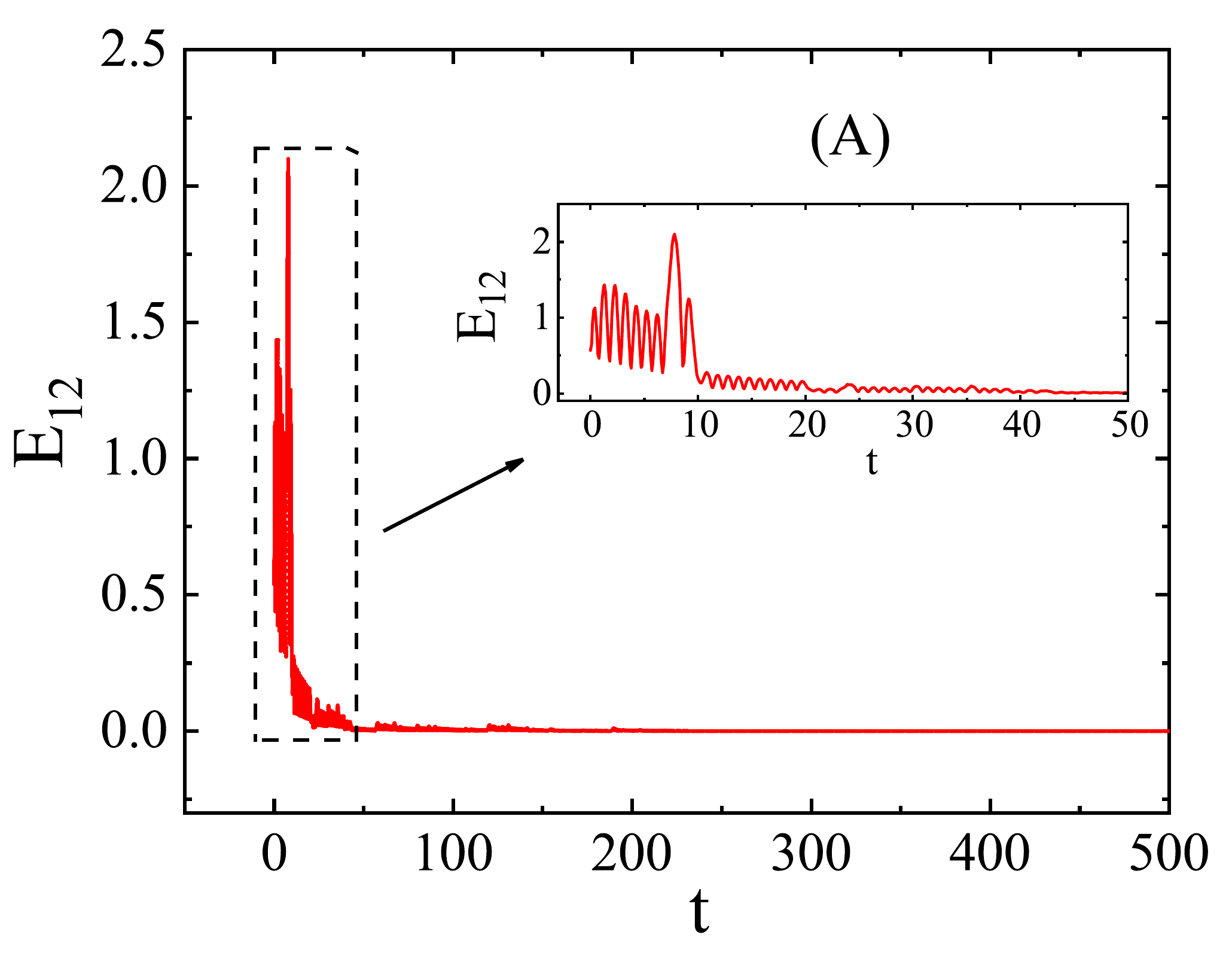}
				\includegraphics[width=61mm]{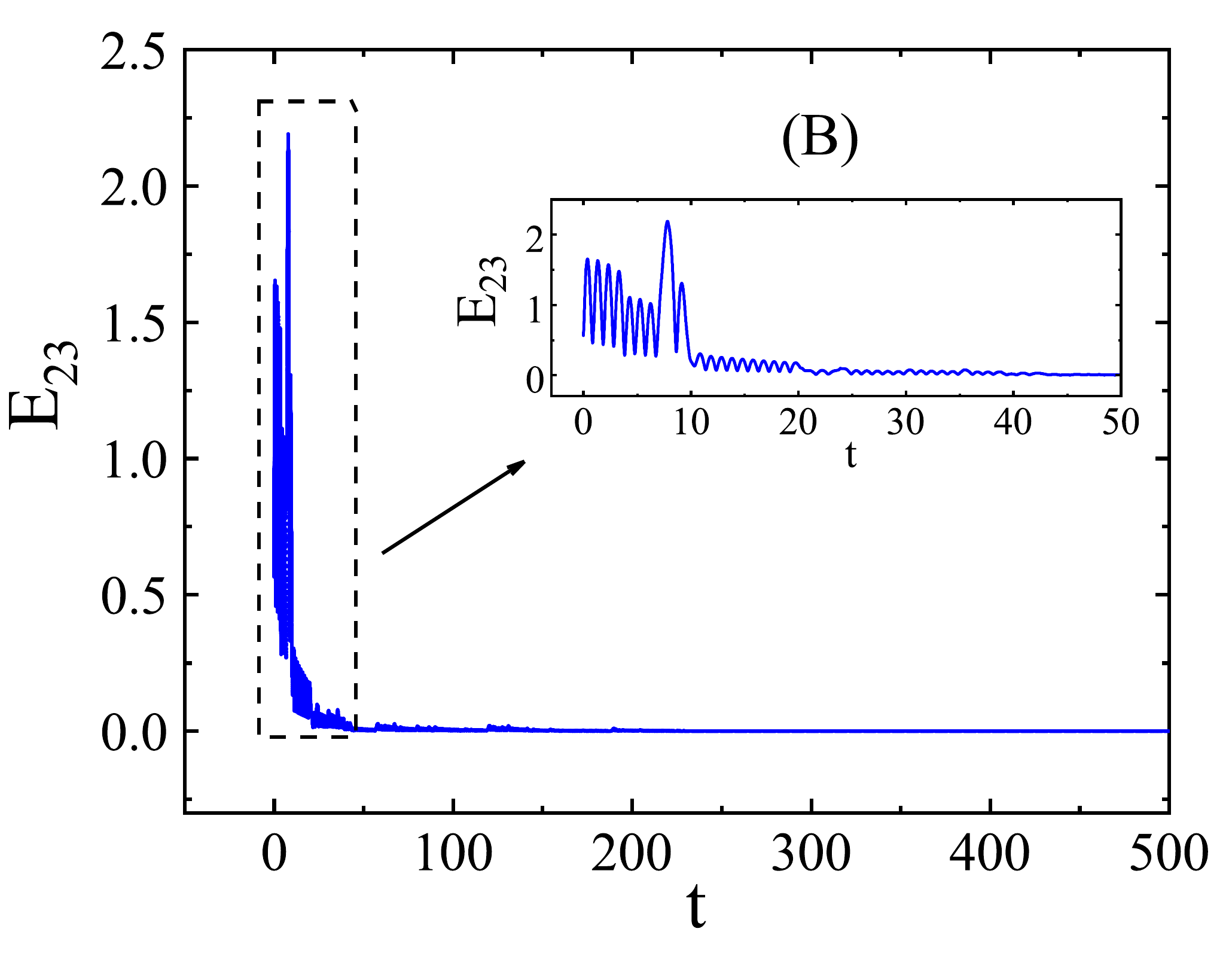}
				\includegraphics[width=61mm]{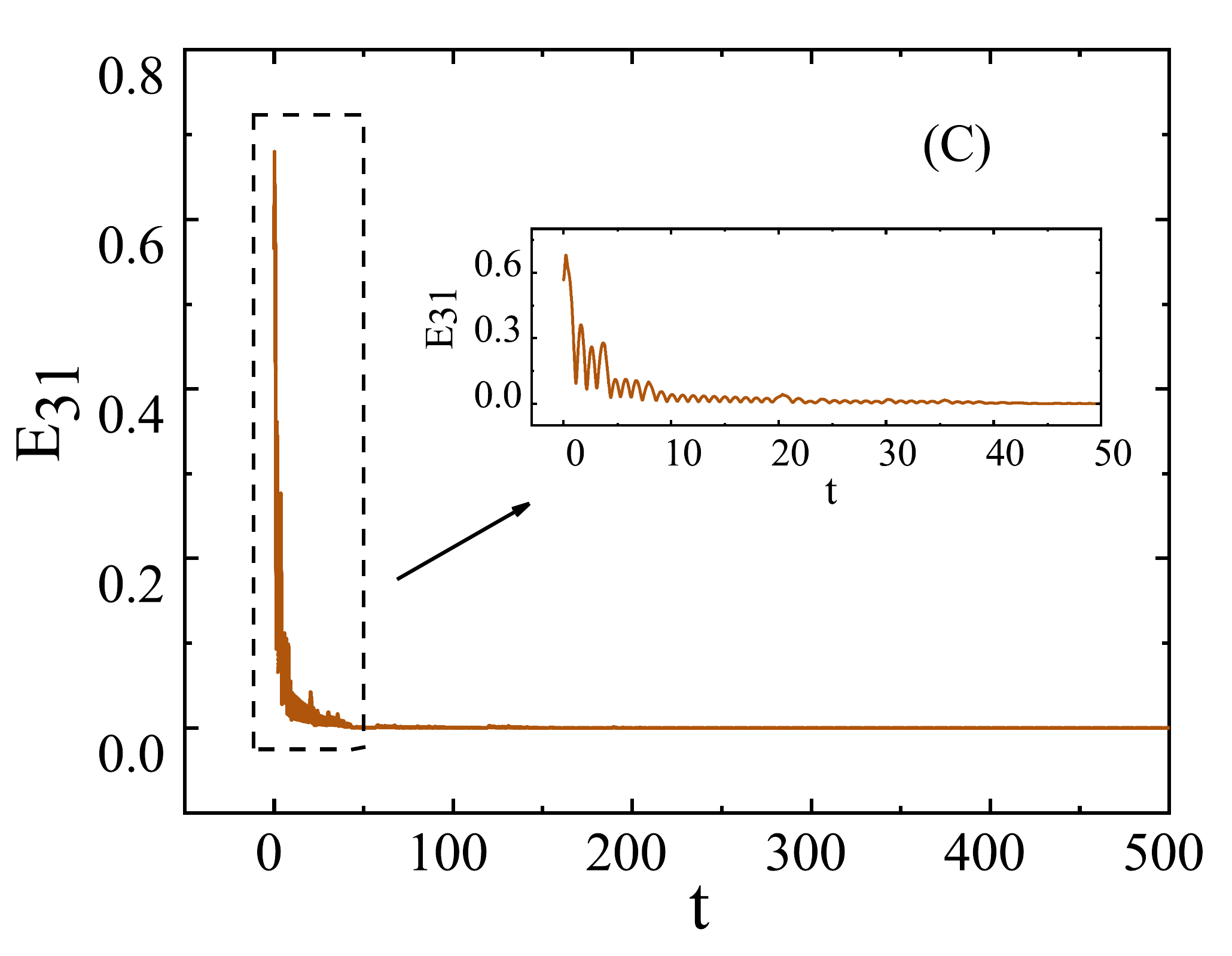}\\
				\includegraphics[width=61mm]{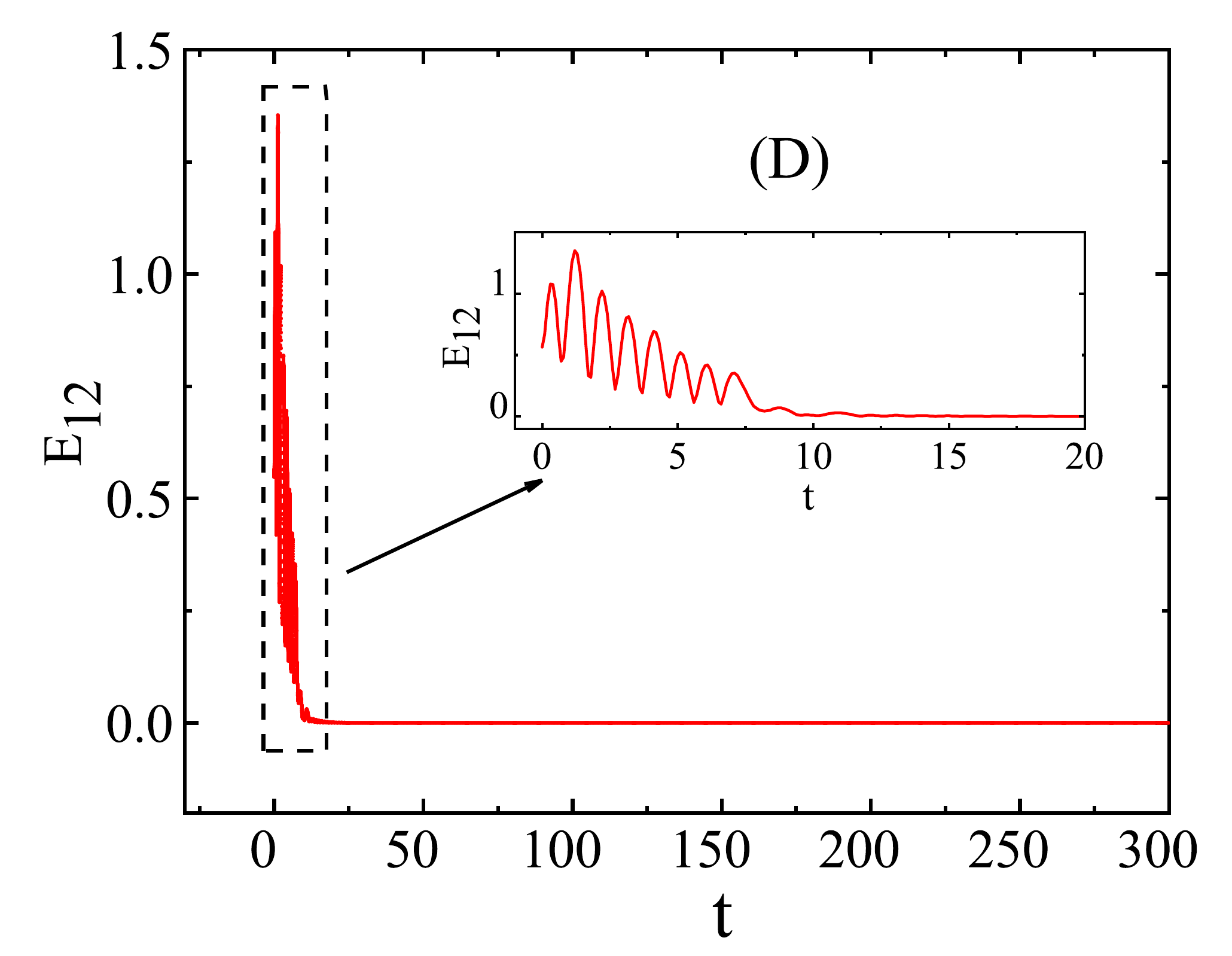}
				\includegraphics[width=61mm]{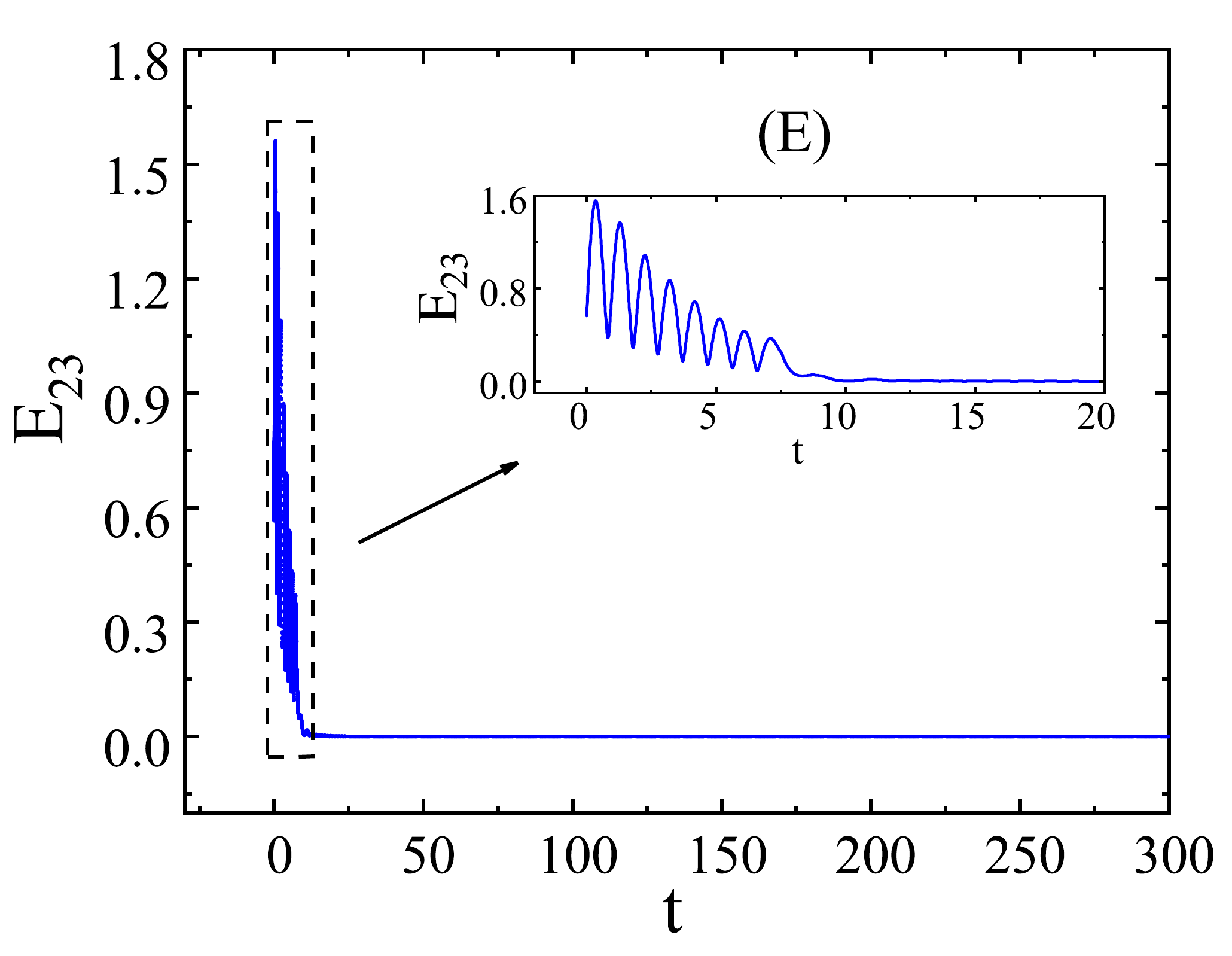}
				\includegraphics[width=61mm]{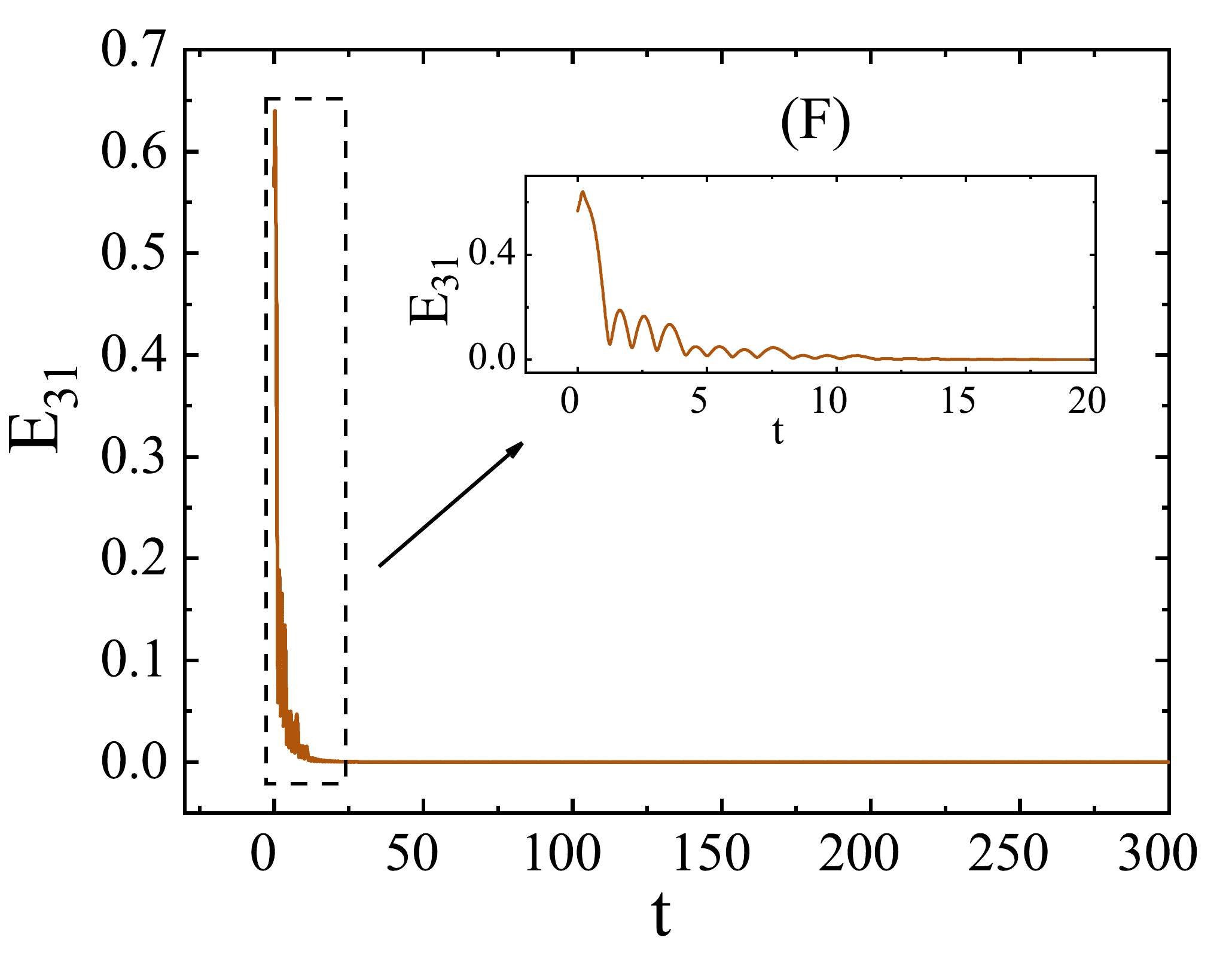}
			\end{tabular}
			\caption{(A),(B),(C) is the evolution of synchronization error between circuits at $\delta_1^C=\delta_2^C=\delta_3^C=0.2$; (D),(E),(F) is the evolution of synchronization error between circuits at$\delta_1^C=\delta_2^C=\delta_3^C=0.4$
				.}
			\label{fig:6}
		\end{figure}
	\end{widetext}

	It can be seen that the closed-loop coupling of capacitor  with a capacitance value of $1.25 $nF ($\delta_3^C=0.1$) does not allow the three Chua circuits to be fully synchronized. However, when we set the coupling strengths $\delta_1^C=\delta_2^C=\delta_3^C=0.2$ ($C_{c_1}=C_{c_2}=C_{c_3}=3.33 $ nF) and $\delta_1^C=\delta_2^C=\delta_3^C=0.4$ ($C_{c_1}=C_{c_2}=C_{c_3}=20 $ nF ), the error evolution is shown in Fig. \ref{fig:6}.
	Eventually the error function between each two circuits tends to zero. At this point we then consider that the three Chua circuits can be fully synchronized at coupling strengths $\delta^C=0.2$ and $\delta^C=0.4$.
	\onecolumngrid

\begin{figure}[ht]
		\centering
		\includegraphics[width=80mm]{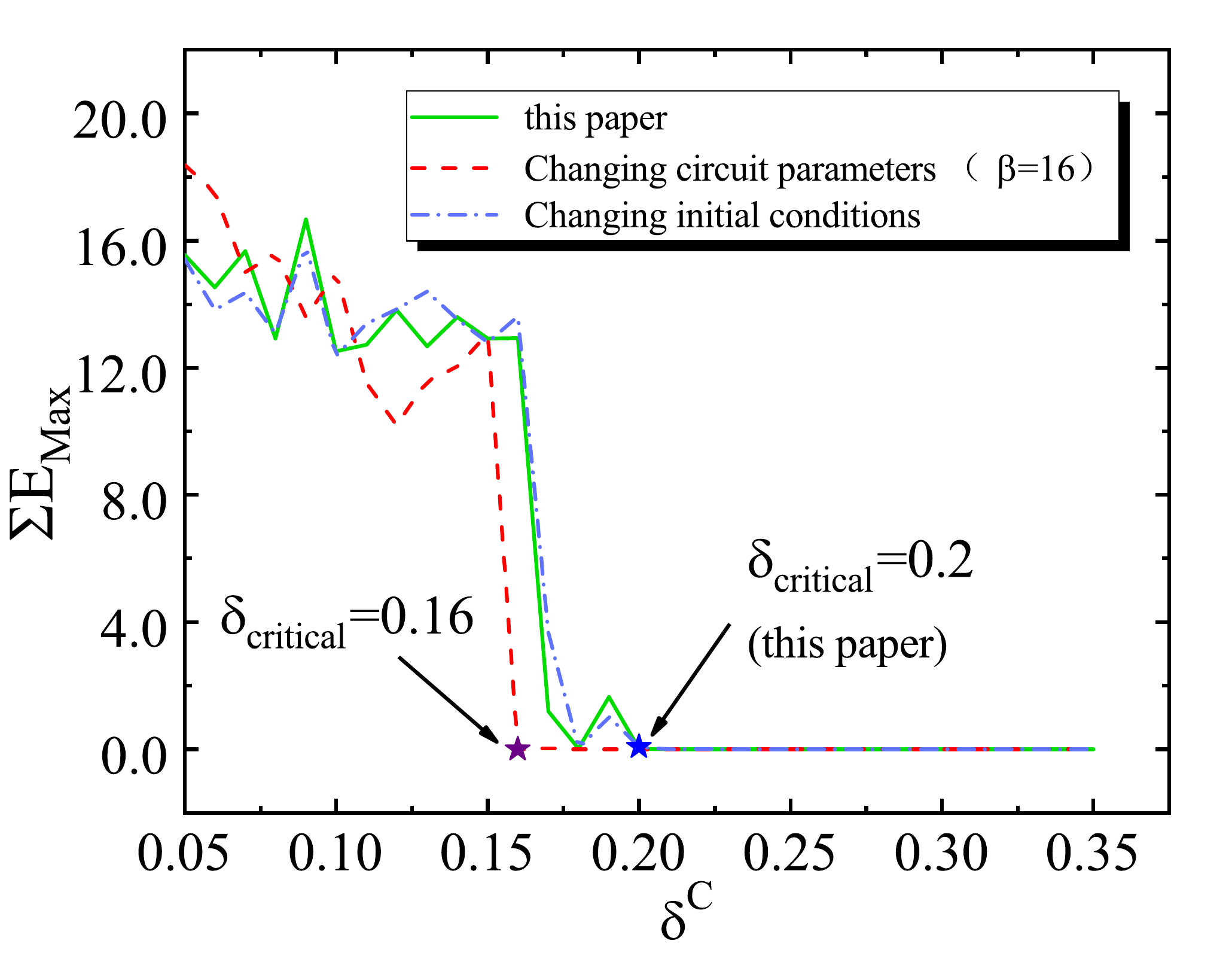}
		\caption{The relationship between the maximum error $\sum E_{Max}$ and the coupling strength $\delta_{critical}^C$ for a system in a given time series. The green solid line corresponds to the parameter [\ref{equ:7}] of the Chua circuit chosen in this paper ; The red dashed line corresponds to the change of $\beta=14.87$ to $\beta=16$ in the original parameter ; The blue dotted line corresponds to the change of the original initial condition to $[x_1(0),y_1(0),z_1(0)]=[0.5,0.5,0.5]$ .}
		\label{BEMAX}
	\end{figure}

\begin{figure}[ht]
		\centering
		\includegraphics[width=80mm]{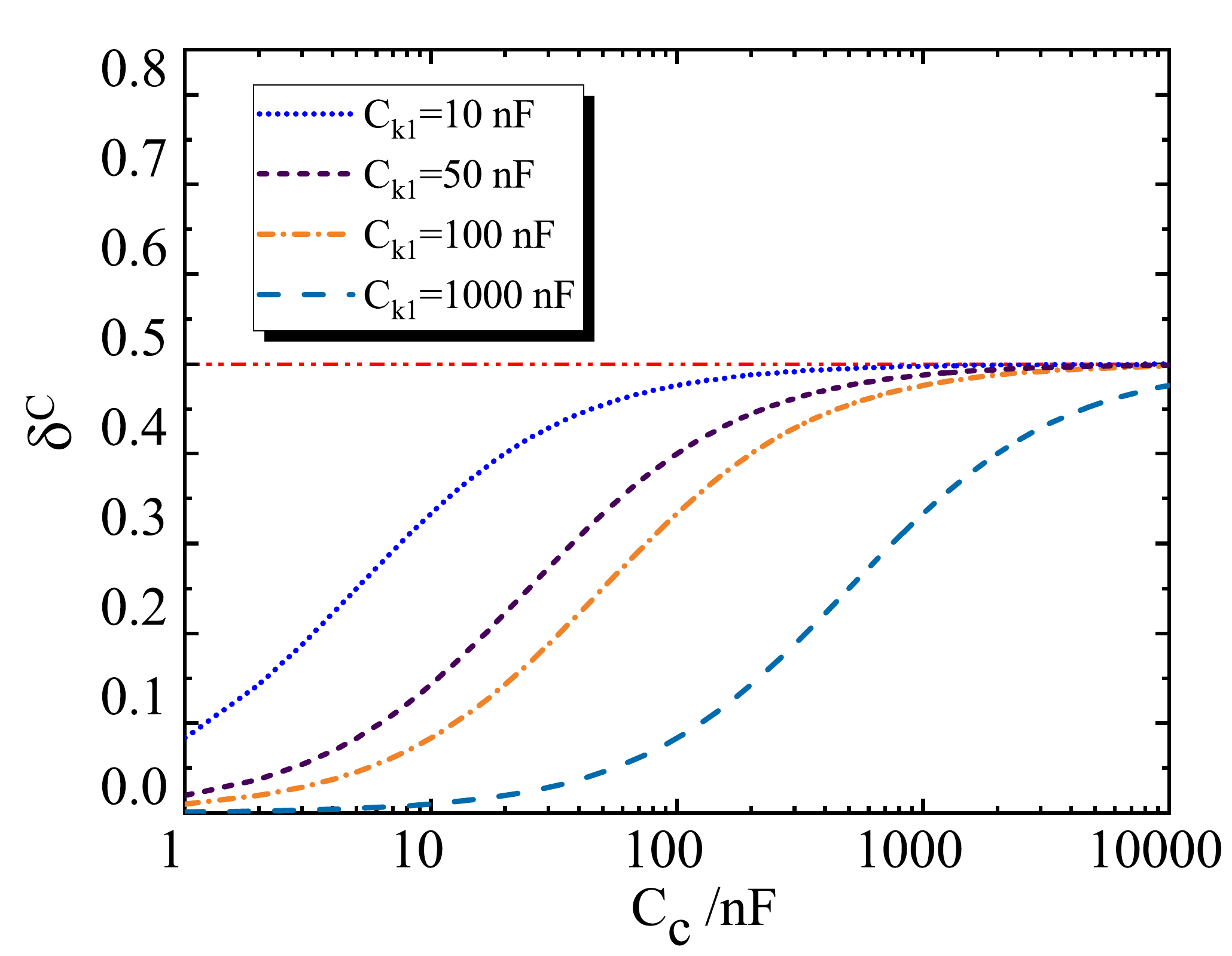}
		\caption{The variation pattern of the capacitive coupling strength with the value of the coupling capacitance. It is noted that, regardless of the value of the capacitance $C_{k1}$ at the output, as the value of the coupling capacitance $C_c$ increases, the maximum value of the coupling strength is $\delta_{max}^C=0.5$ .}
		\label{limit0.5}
	\end{figure}

	However, is there a threshold coupling strength that can exactly synchronize the three Chua circuits coupled through the closed loop of capacitance? The answer is yes. We calculated the maximum sum of errors $\sum E_{Max}$ of the system in a certain time series and found this threshold coupling strength $\delta_{critical}^C$ (Fig. \ref{BEMAX}). It was also found that the value of this threshold coupling strength varies with the change of circuit parameters and initial conditions. That is, for different Chua circuits, the threshold coupling strength that will enable them to be fully synchronized will be different. In this paper, we choose a Chua circuit corresponding to a threshold coupling strength $\delta_{critical}^C=0.2$. However, for capacitive coupling, the upper limit of the coupling strength is limited by Eq. (\ref{equ:11}). The maximum value of the coupling strength for capacitive coupling is $\delta_{max}^C=\lim_{{C_c\rightarrow\infty}}=0.5$ (Fig. \ref{limit0.5}) .Therefore, when $\delta^C \in [\delta_{critical}^C , \delta_{max}^C]$ is selected, the three Chua circuits can be completely synchronized; when $\delta^C \in [0 ,\delta_{critical}^C]$ is selected, the three Chua are not completely synchronized.

	\begin{figure*}[ht]
		\centering
		\begin{tabular}{ccc}
			\includegraphics[width=61mm]{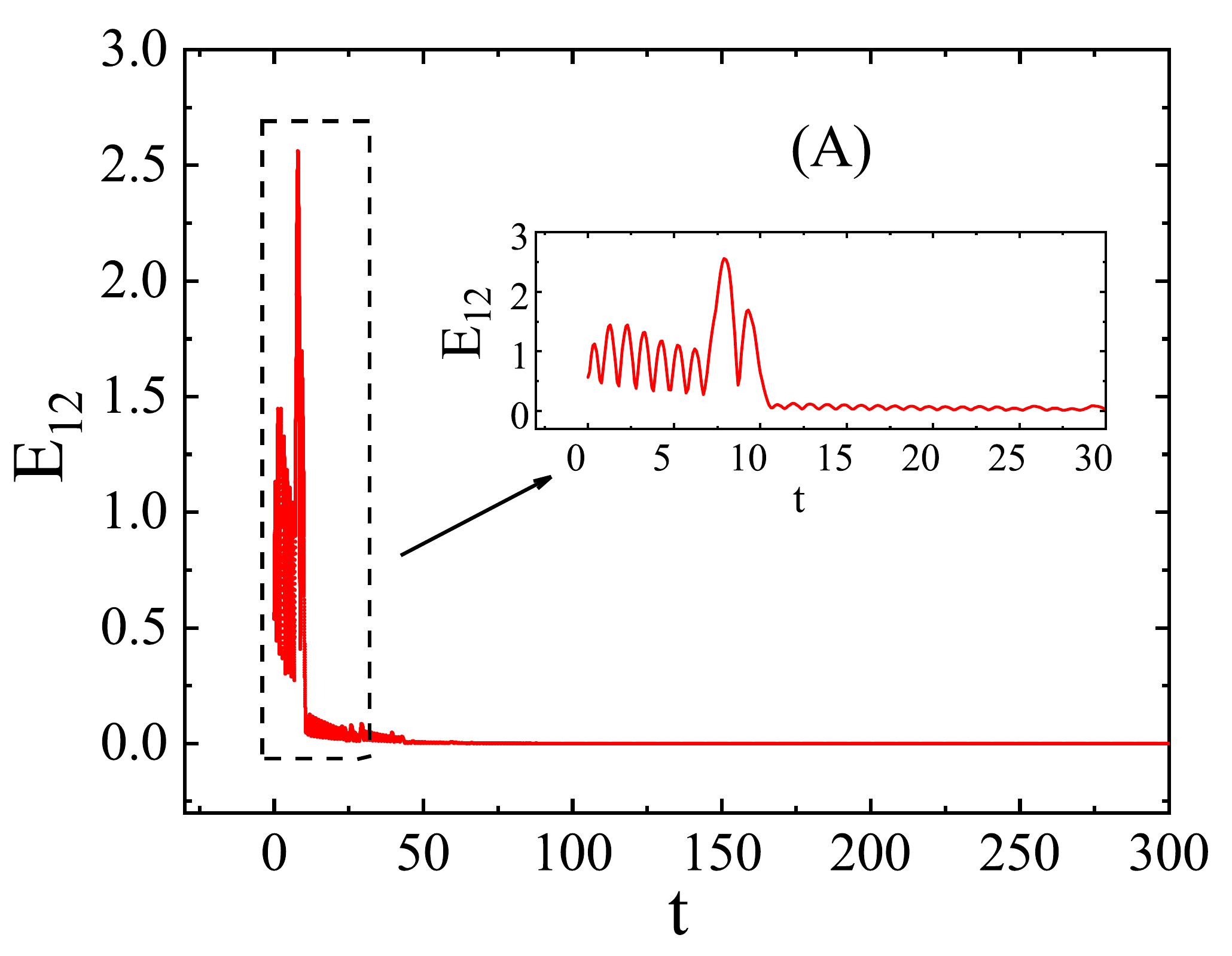}
			\includegraphics[width=61mm]{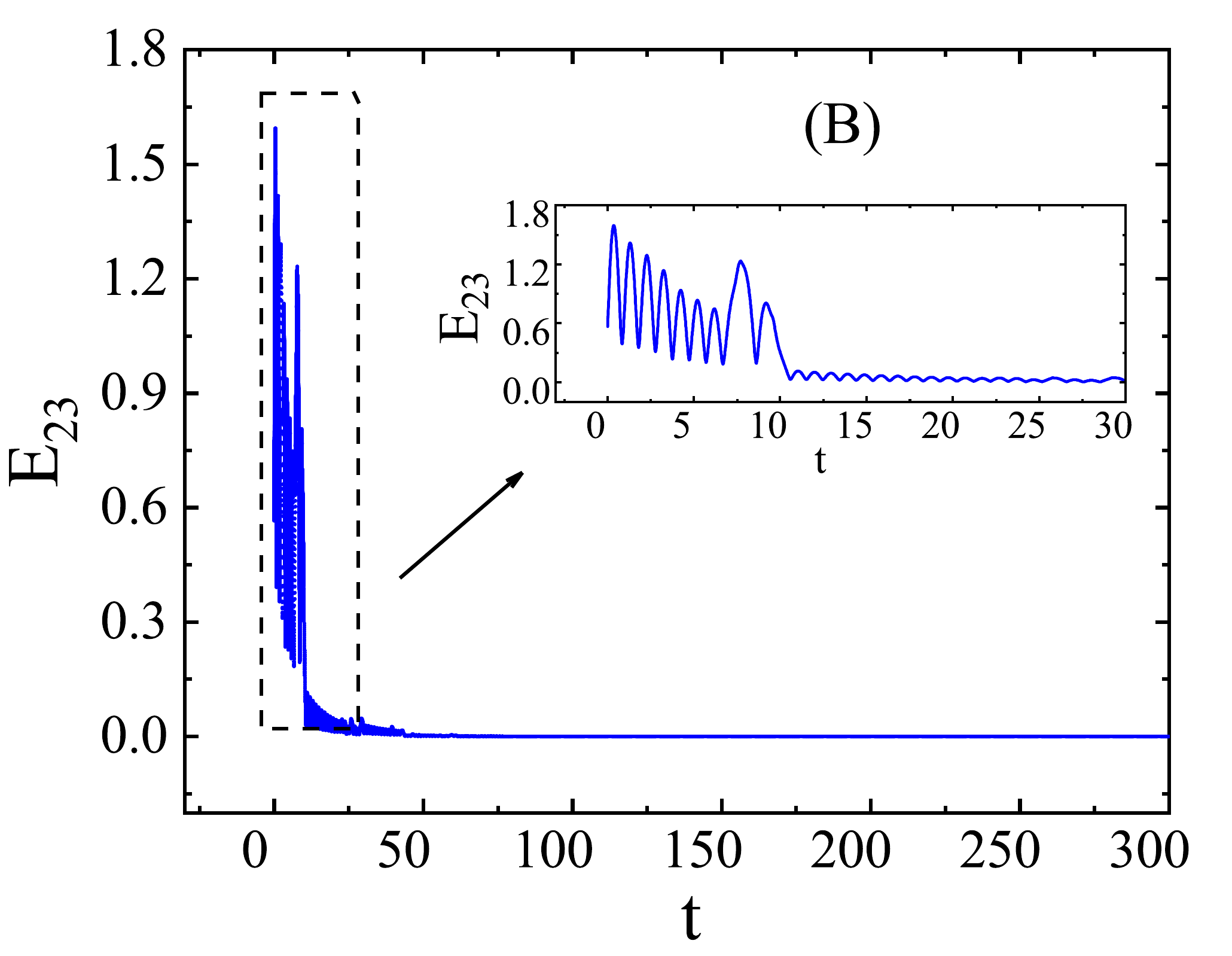}
			\includegraphics[width=61mm]{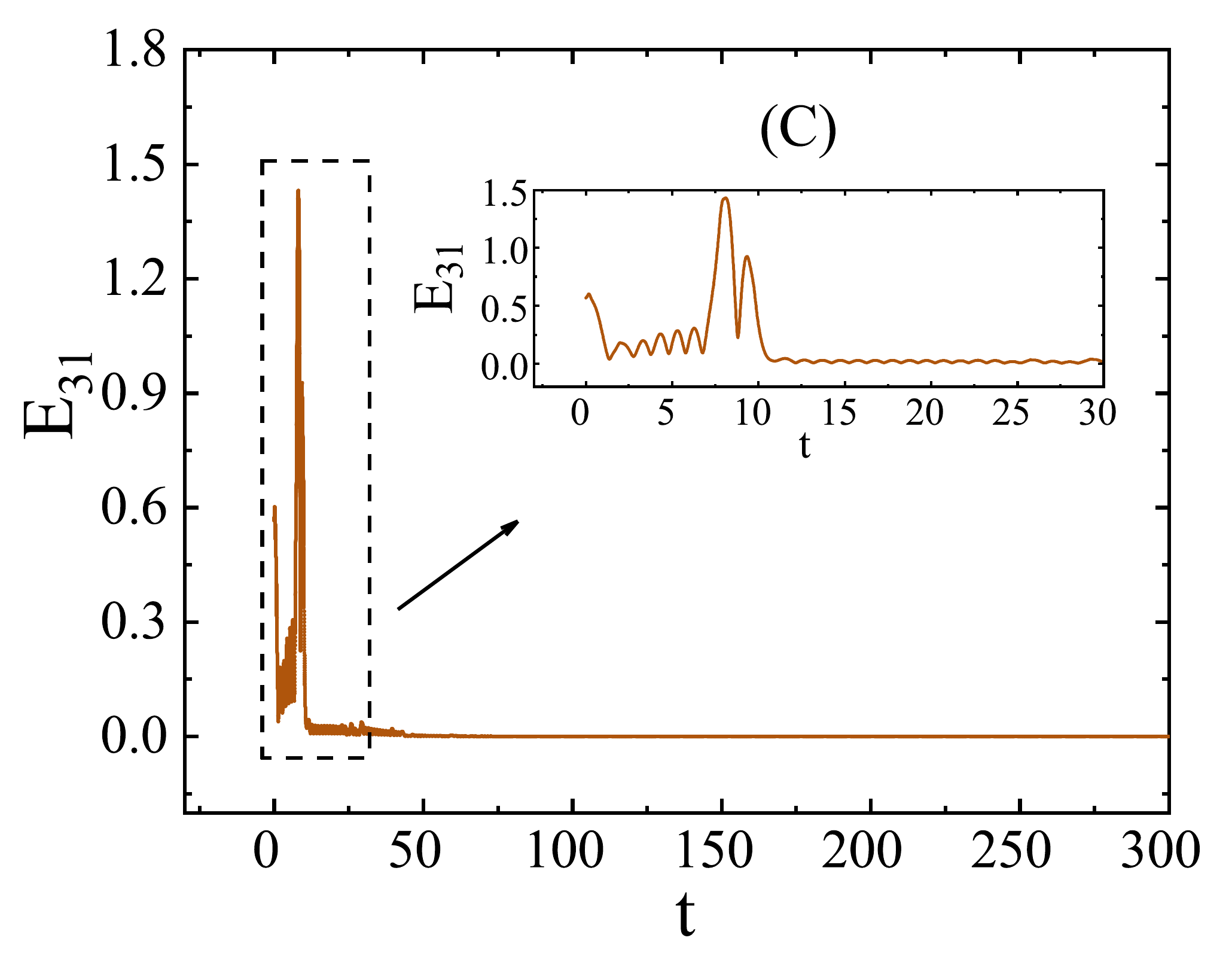}
		\end{tabular}
		\caption{Special result for capacitive closed-loop coupling: there is a coupling strength that does not reach the synchronizable coupling strength threshold $\delta_{critical}^C$ (The coupling strengths are $[\delta_1^C , \delta_2^C , \delta_3^C]=[0.1,0.4,0.45]$, and $\delta_1^C=0.1<\delta_{critical}^C$ ) . }
		\label{fig:7}
	\end{figure*}

	In fact, when we couple three Chua circuits by closed-loop capacitance. If one of the coupling strengths does not reach the synchronization threshold, we can also make the three circuits fully synchronized by appropriately adjusting the other coupling strength parameters. For example, we set the coupling strength to $[\delta_1^C , \delta_2^C , \delta_3^C]=[0.1,0.4,0.45]$, where $\delta_1^C<\delta_{critical}^C$. However, the error evolution diagram (Fig. \ref{fig:7}) shows that the three Chua circuits are eventually synchronized in the same short time. This is something that cannot be done with unidirectional coupling.

	\subsection{Closed-loop Coupling Advantage Analysis}
	We consider three Chua circuits with the same circuit parameters and different initial conditions (same parameters and initial states as discussed in Sec. \ref{sec:3}). All three Chua circuits are in a chaotic state and can excite a double vortex attractor. We use capacitors  for one-way and closed-loop coupling of the three Chua circuits,  and discuss the synchronization of the three circuits under different coupling methods.For the subsequent discussion. The coupling strengths we have chosen are the same, i.e., $\delta_1^C=\delta_2^C=\delta_3^C=\delta^C $.
	
\begin{figure*}[h]
		\centering
		\begin{tabular}{cccccc}
			\includegraphics[width=61mm]{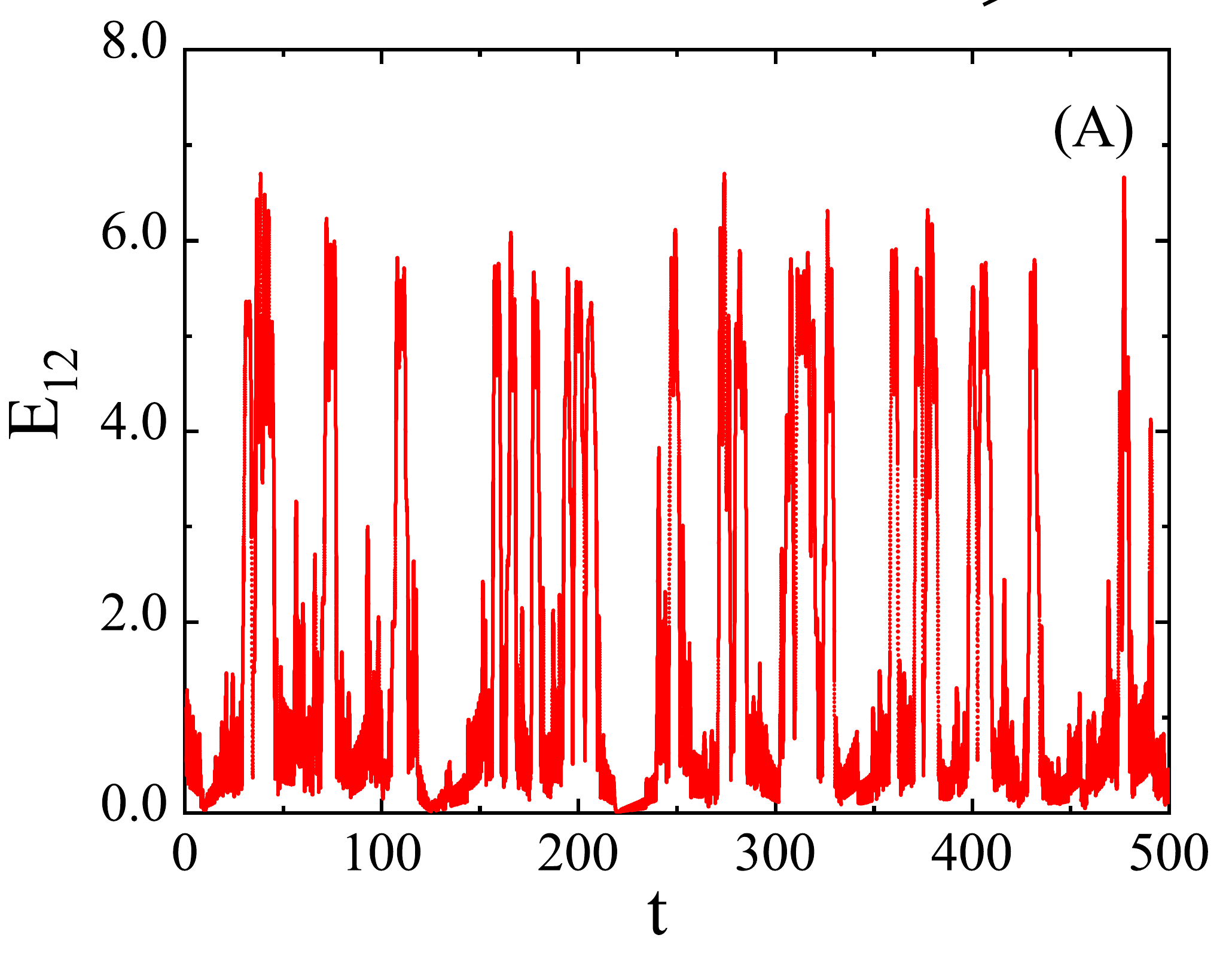}
			\includegraphics[width=61mm]{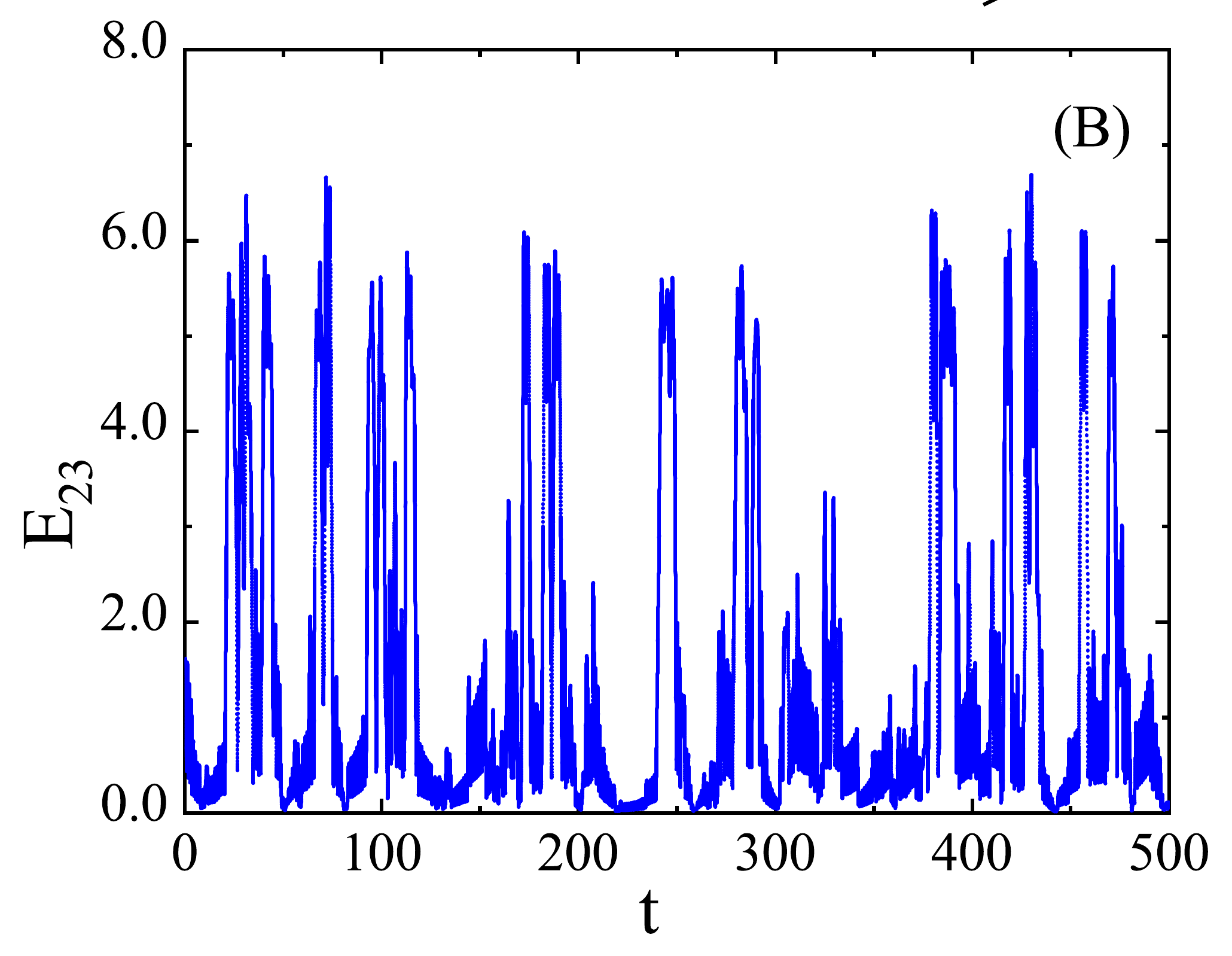}
			\includegraphics[width=61mm]{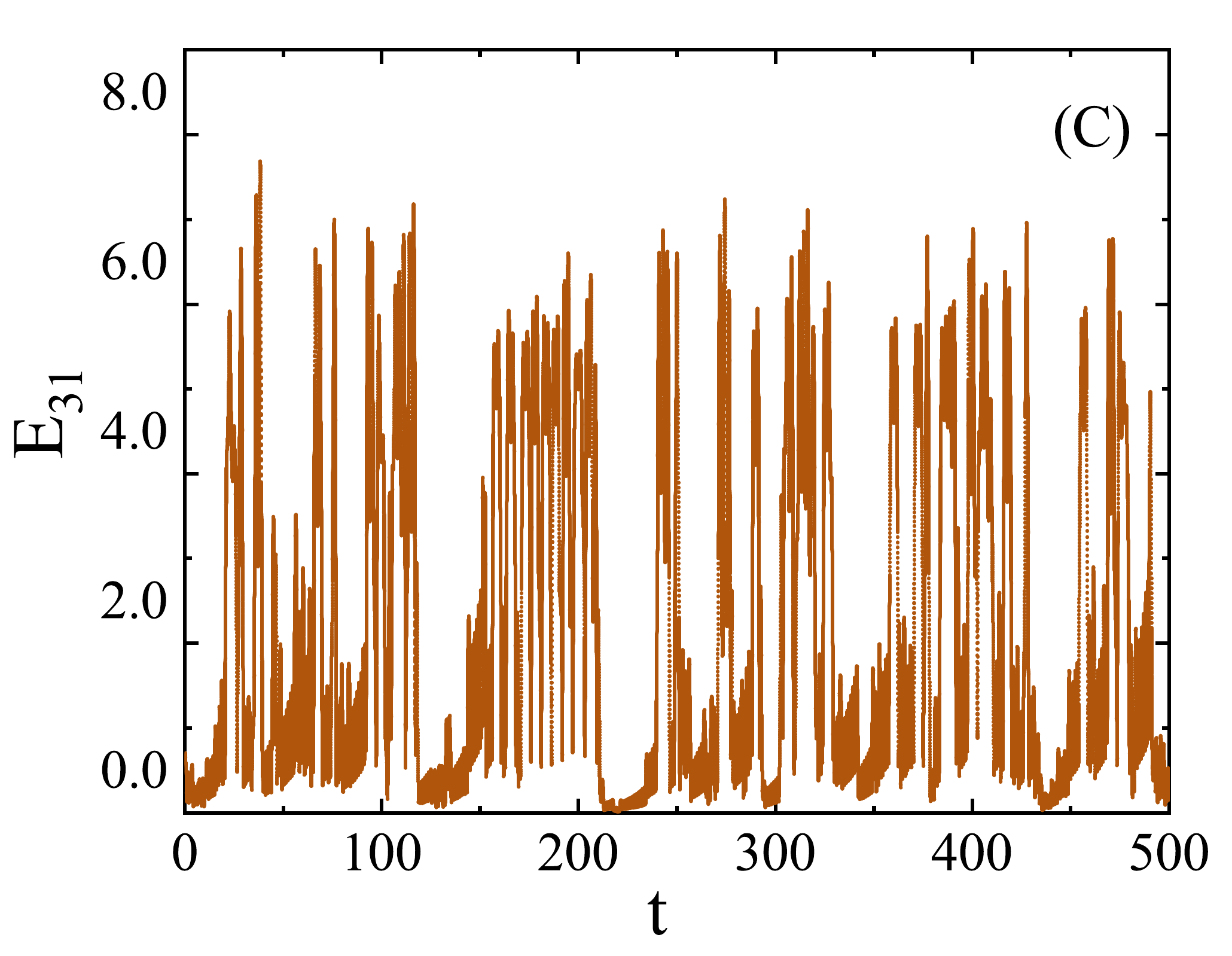}\\
			\includegraphics[width=61mm]{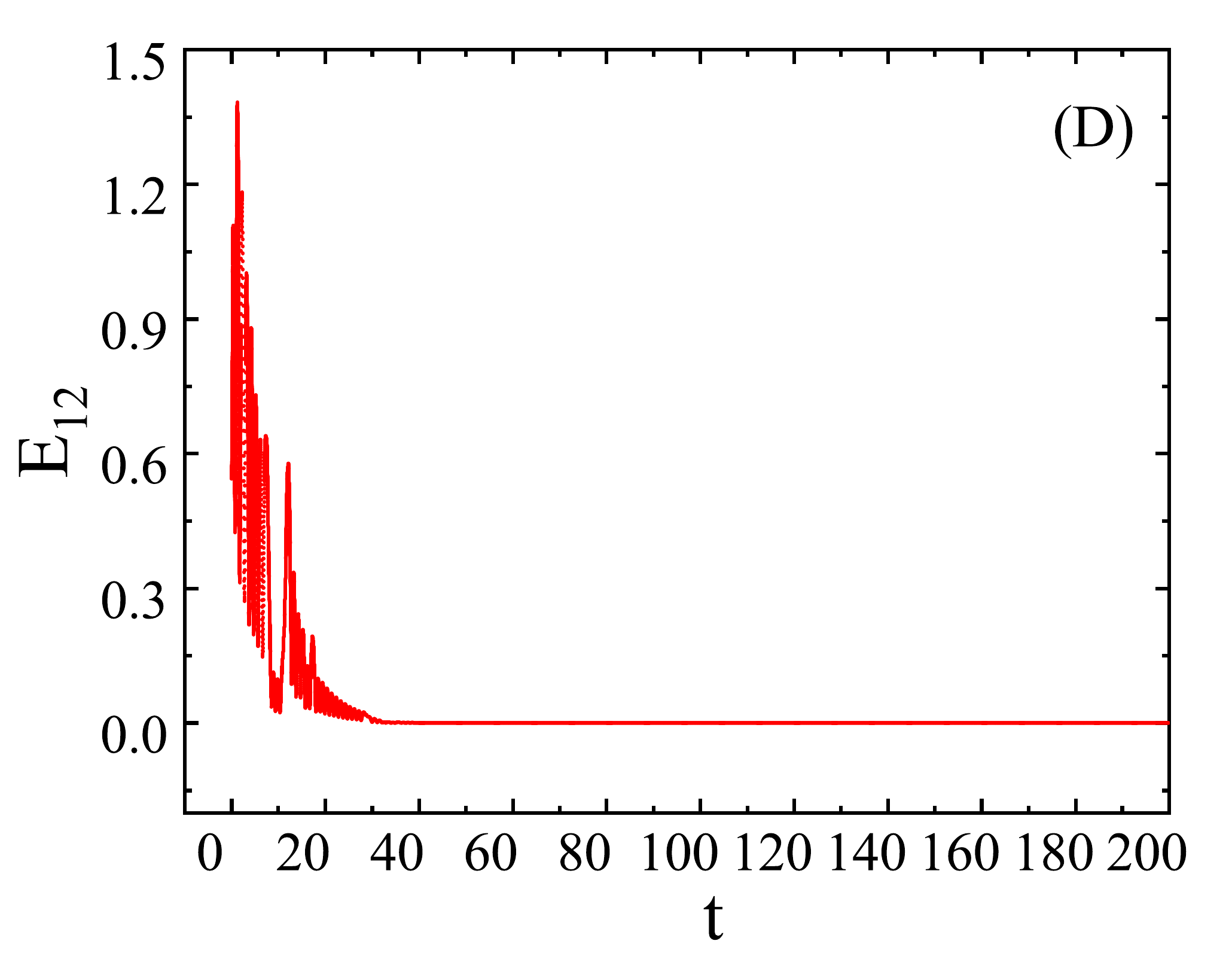}
			\includegraphics[width=61mm]{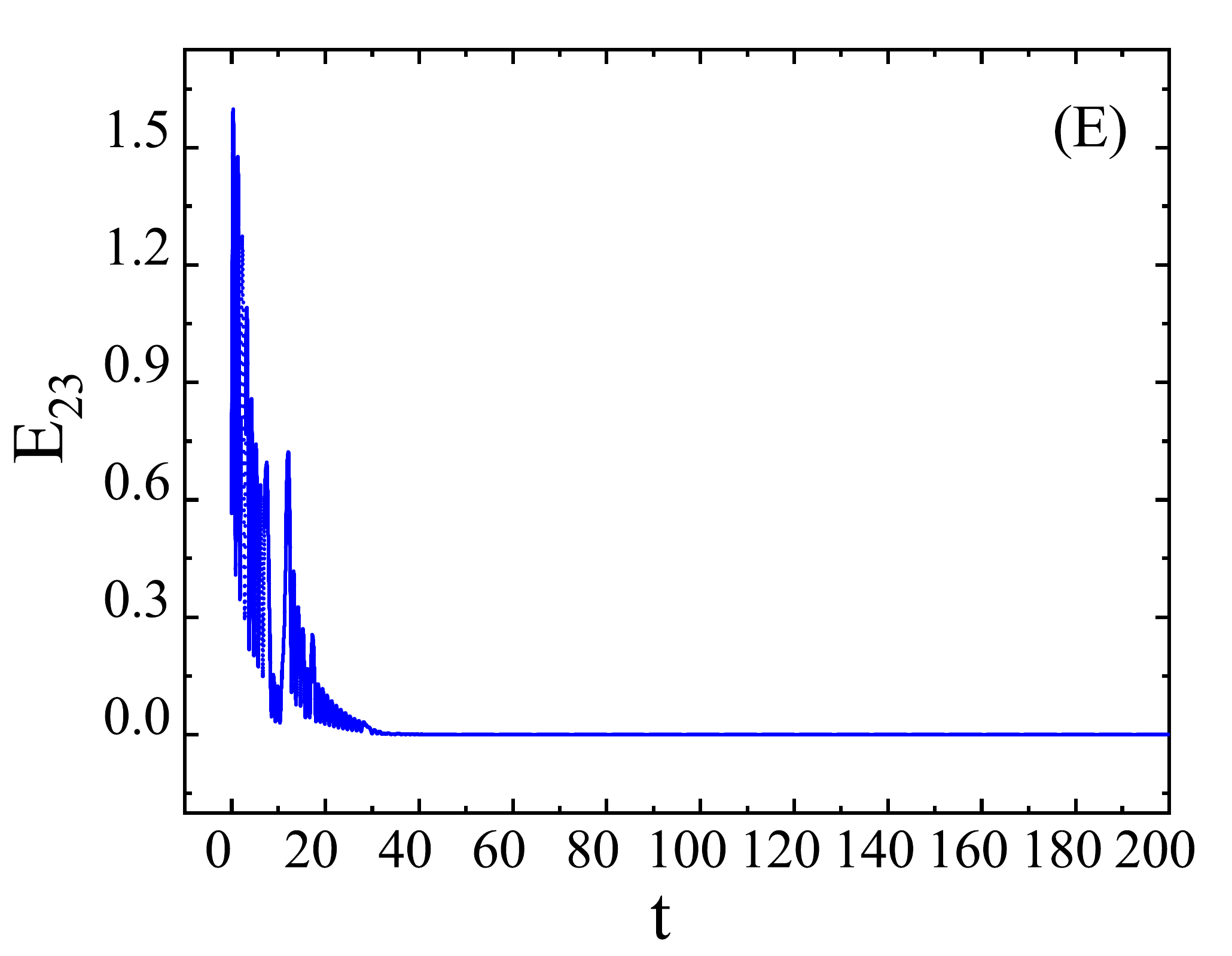}
			\includegraphics[width=61mm]{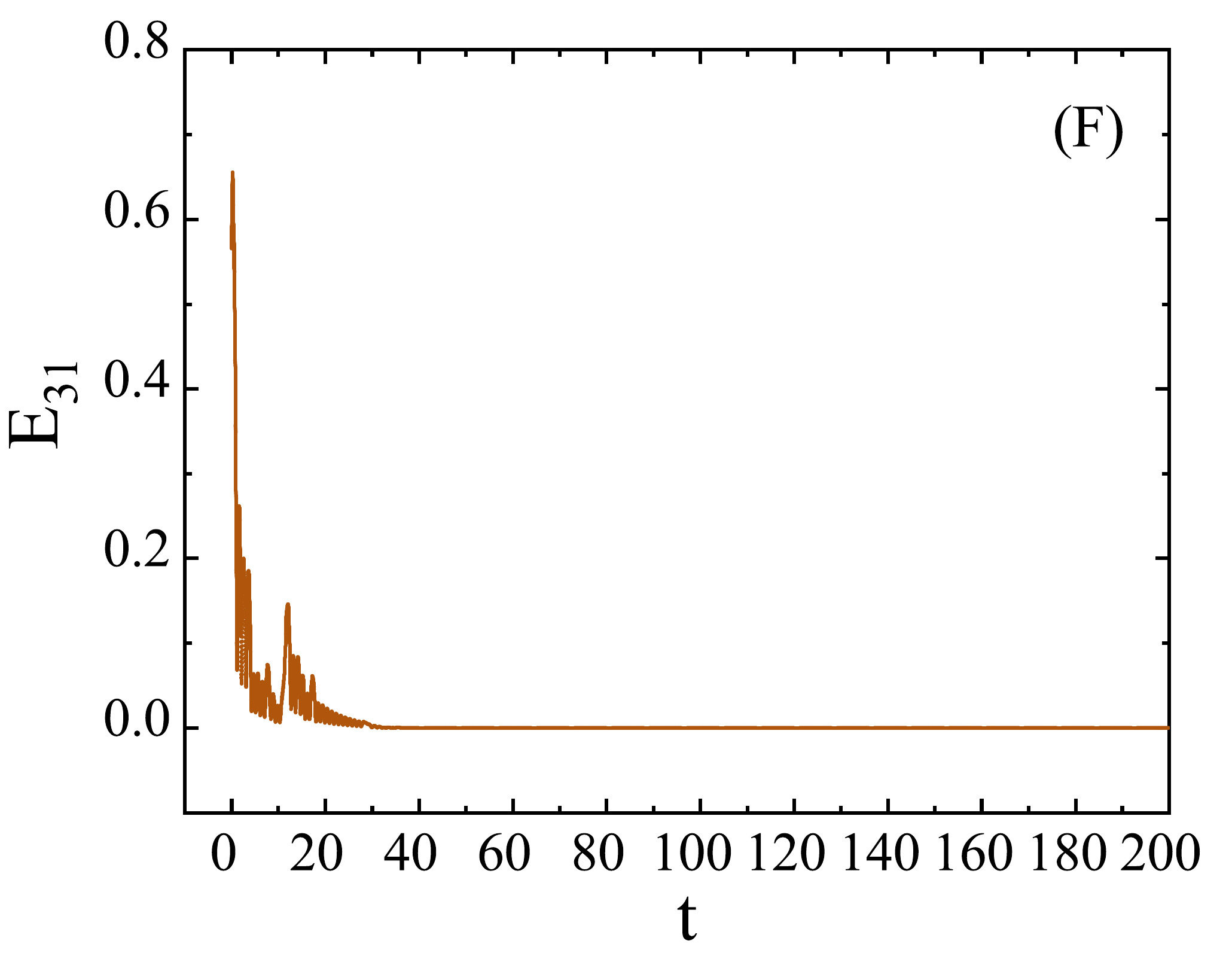}
		\end{tabular}
		\caption{For the same three Chua circuits, at the same coupling strength $\delta^C=0.3$. (A) , (B) , (C) is the synchronization error between circuits when coupled unidirectionally through capacitors; (D) , (E) , (F) is the synchronization error between circuits when coupled closed-loop through capacitors
			.}
		\label{fig:11}
	\end{figure*}

\begin{figure*}[h]
		\centering
		\begin{tabular}{cccccc}
			\includegraphics[width=61mm]{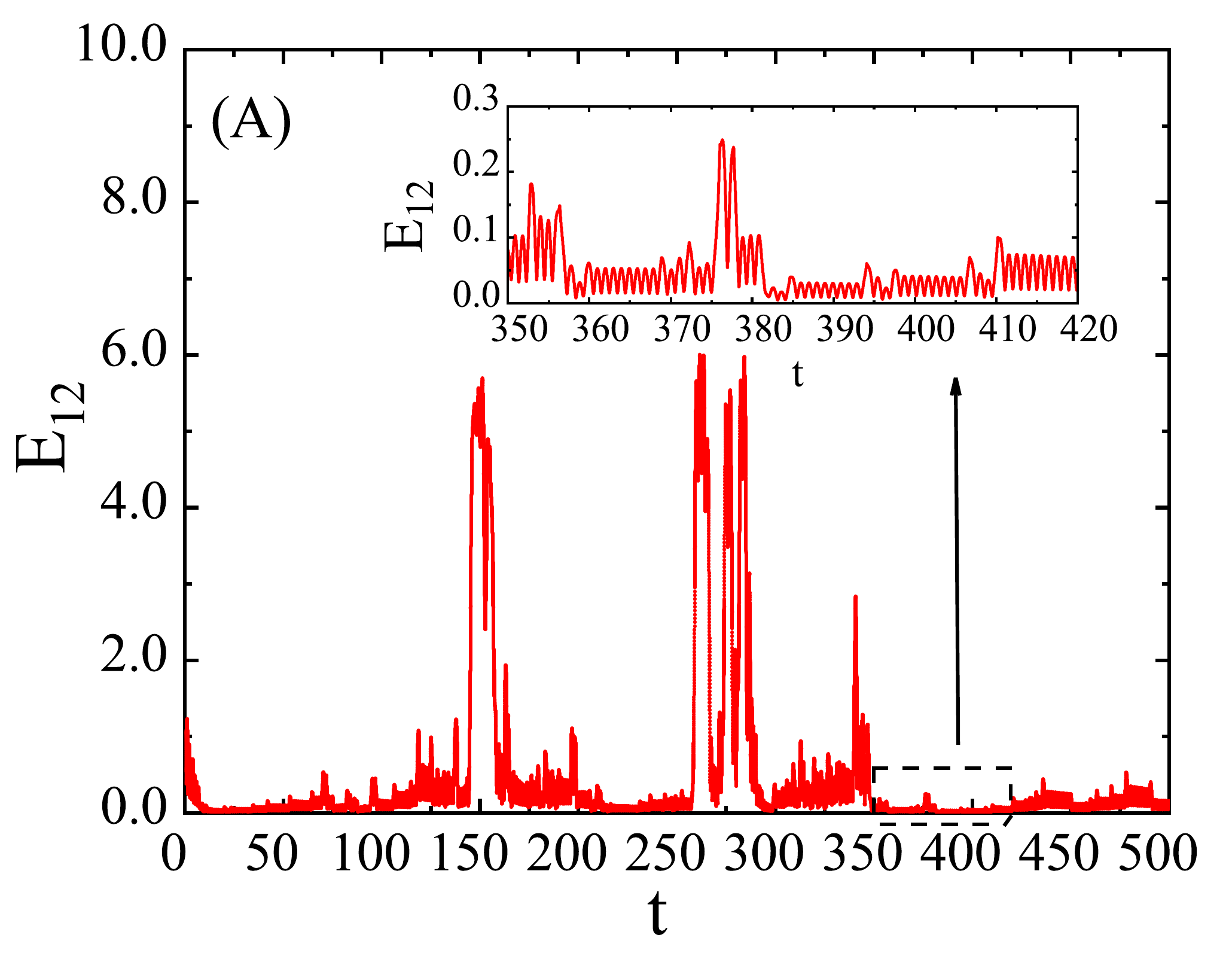}
			\includegraphics[width=61mm]{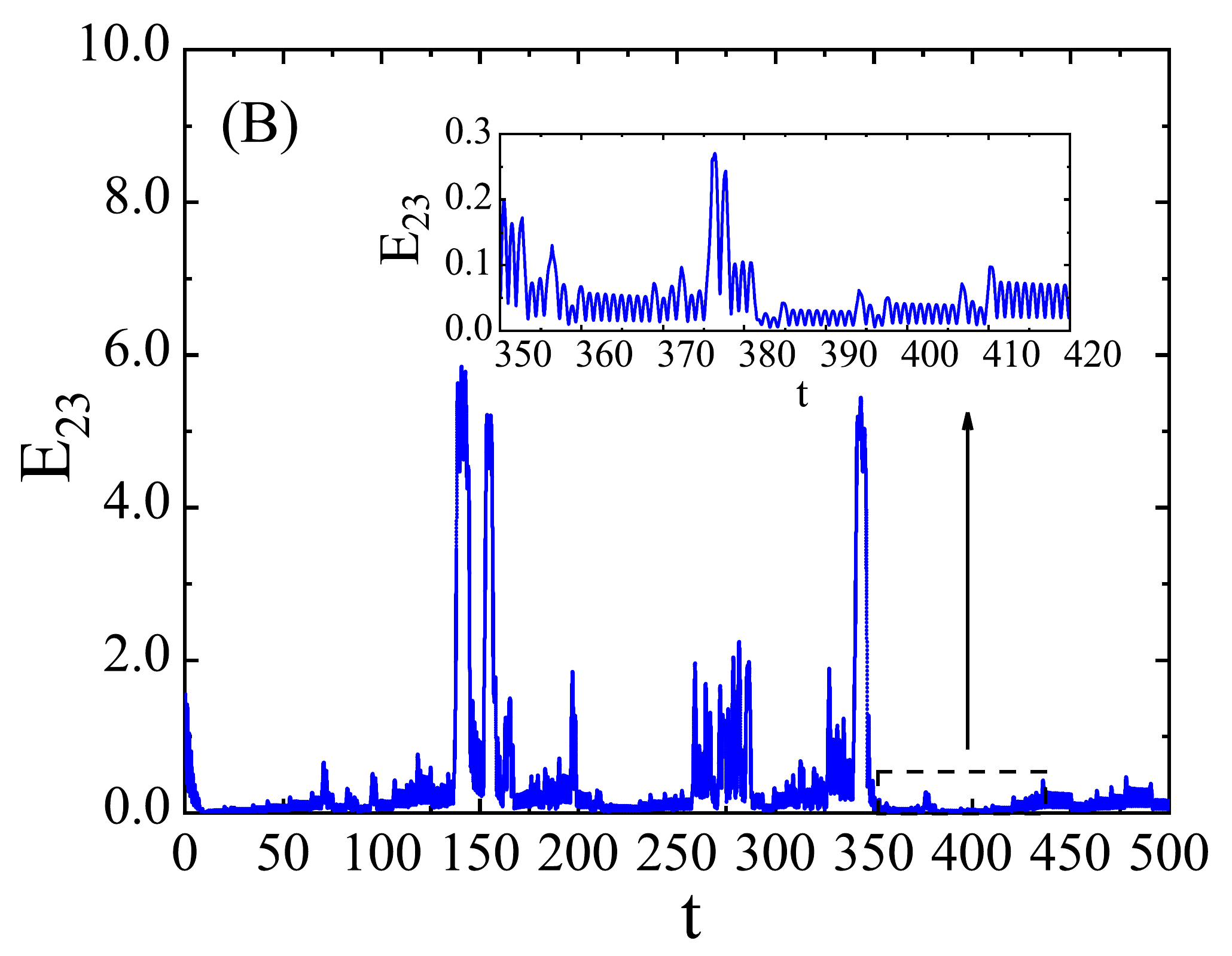}
			\includegraphics[width=61mm]{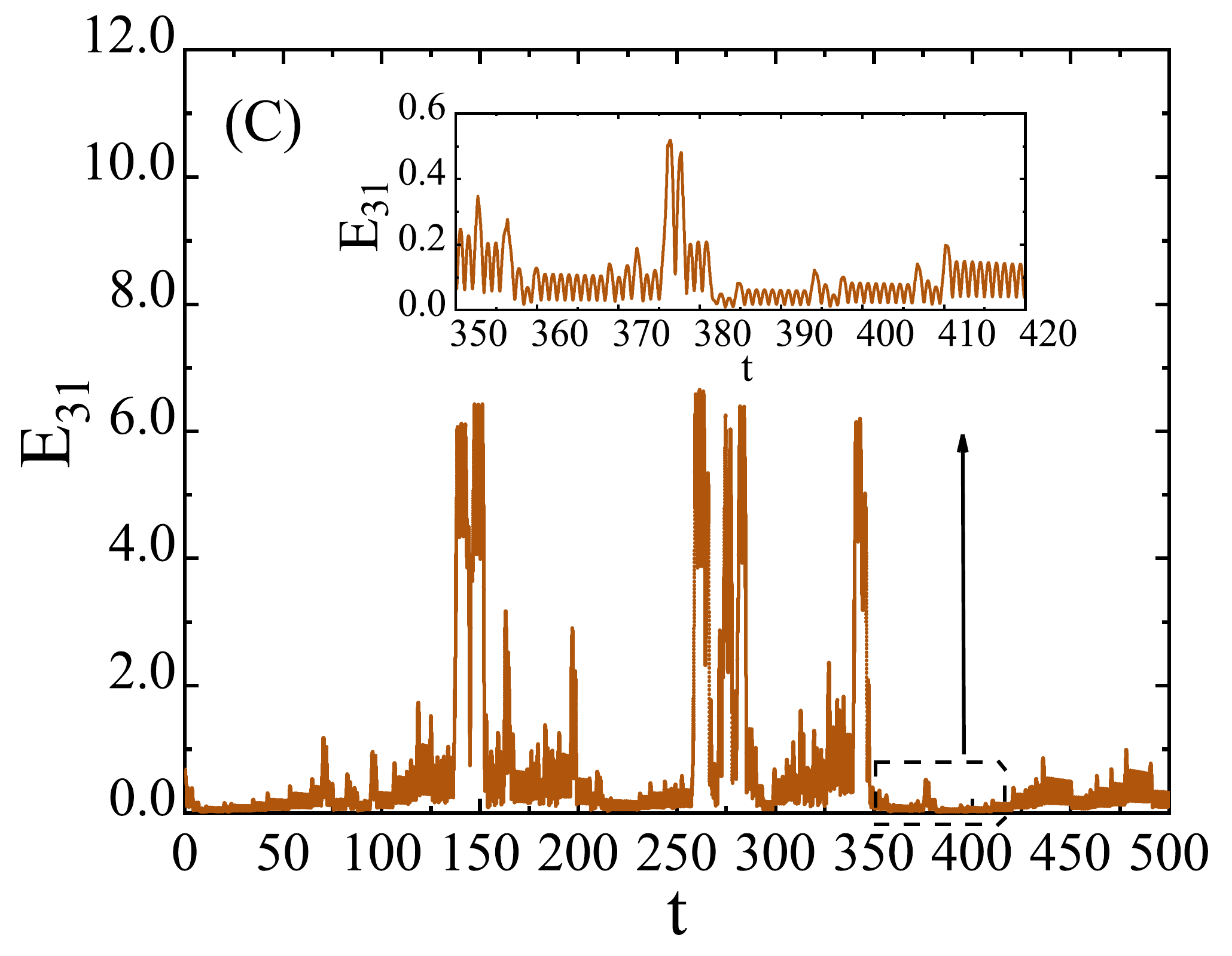}\\
			\includegraphics[width=61mm]{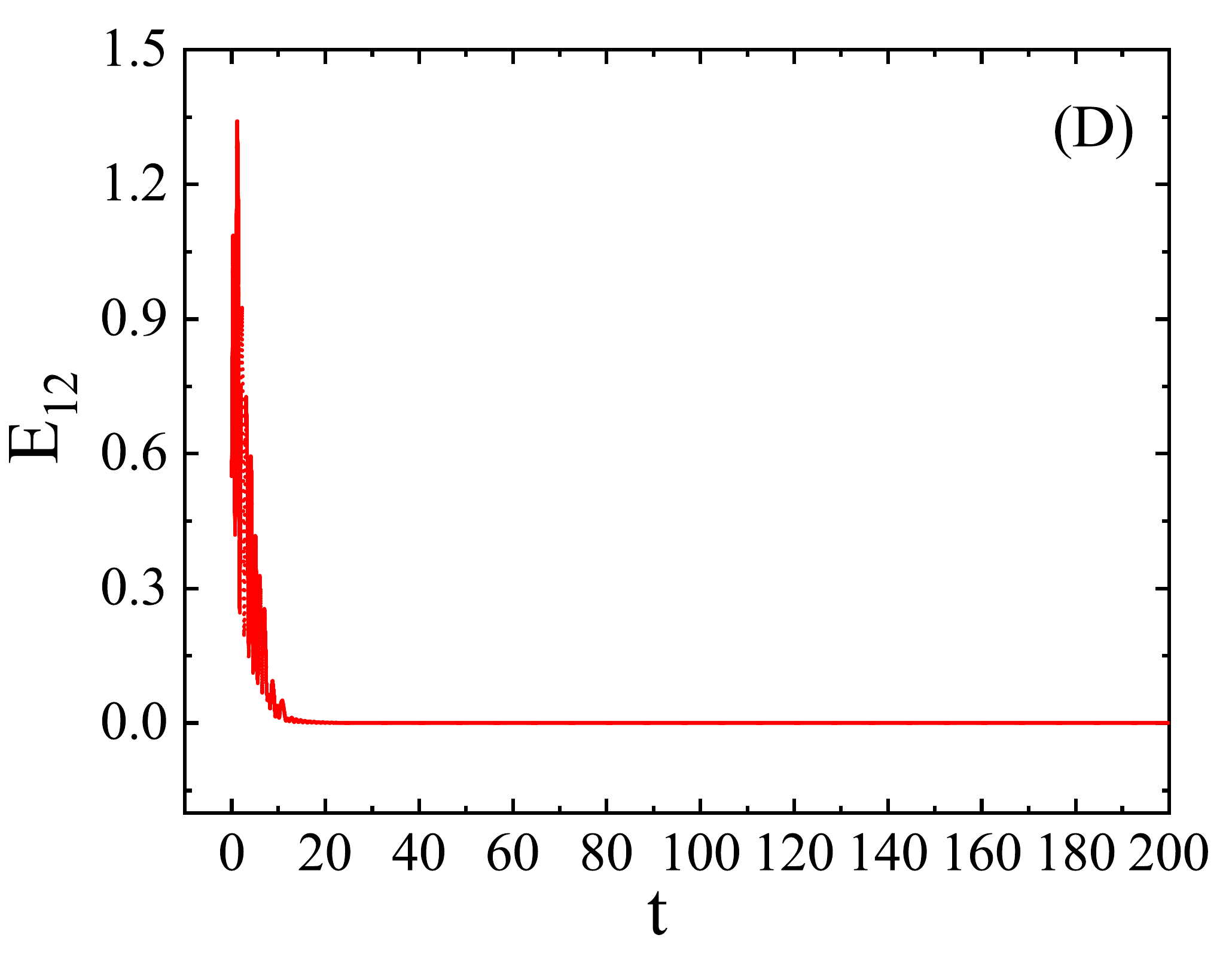}
			\includegraphics[width=61mm]{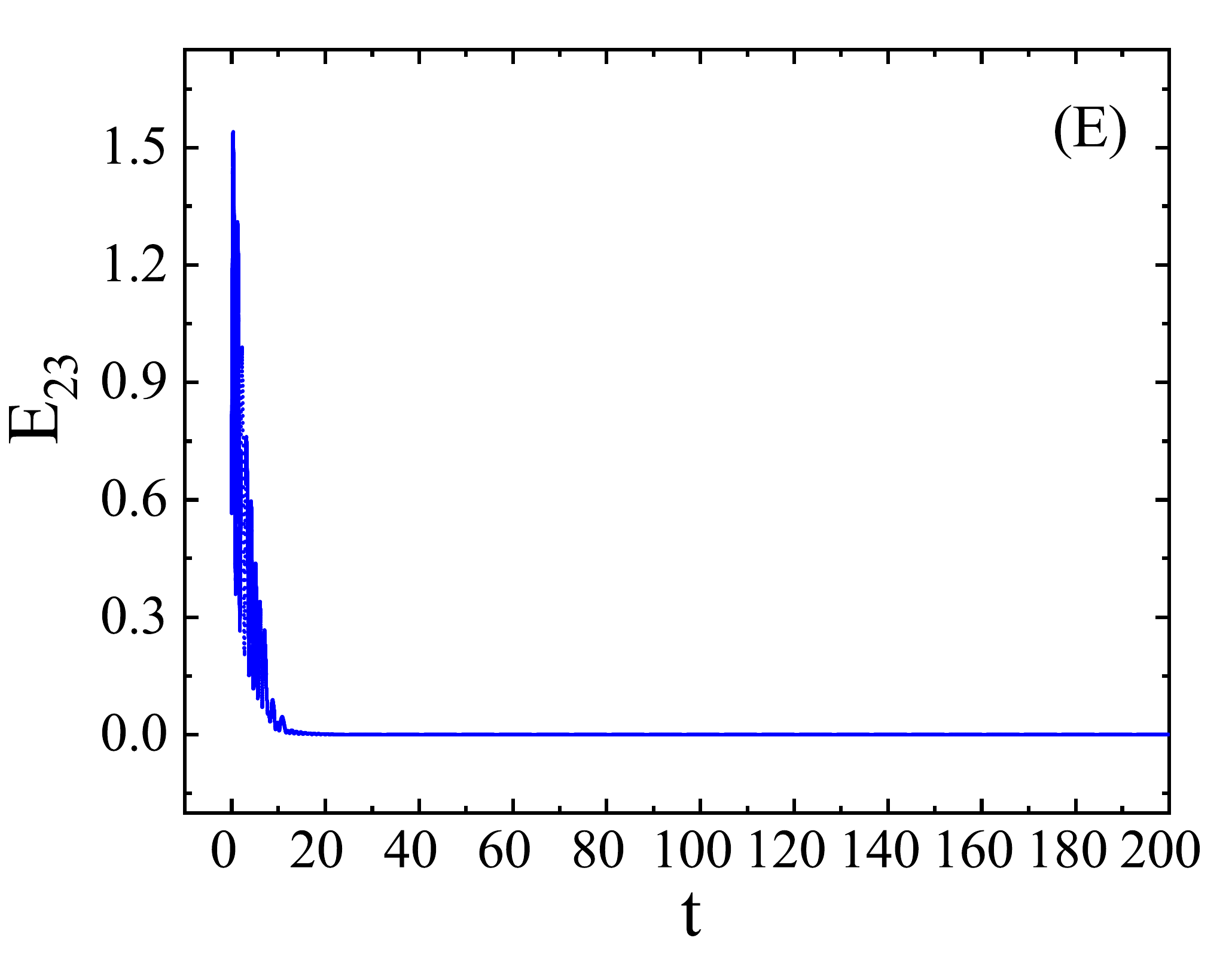}
			\includegraphics[width=61mm]{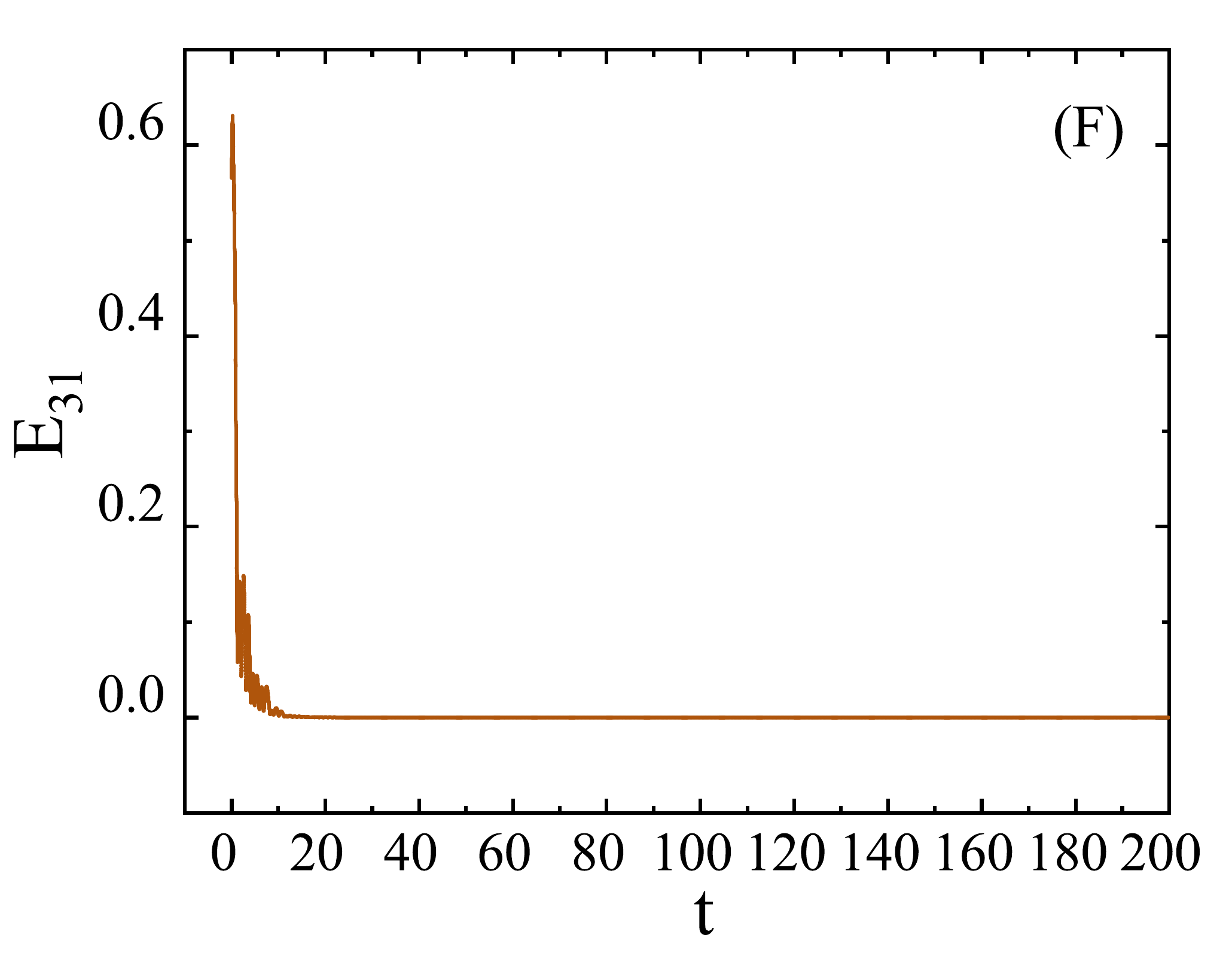}
		\end{tabular}
		\caption{For the same three Chua circuits, at the same coupling strength $\delta^C=0.48$. (A) , (B) , (C) is the synchronization error between circuits when coupled unidirectionally through capacitors; (D) , (E) , (F) is the synchronization error between circuits when coupled closed-loop through capacitors
			.}
		\label{fig:12}
	\end{figure*}

	We  discuss the unidirectional coupling and Closed-loop coupling of the three Chua circuits using capacitors , respectively. We all set the coupling strengths to $\delta^C=0.3$ and $\delta^C=0.48$, respectively, and the circuit coupling error evolution diagram is shown in Fig. \ref{fig:11} and Fig. \ref{fig:12}.
	
\begin{figure}[ht]
		\centering
		\includegraphics[width=80mm]{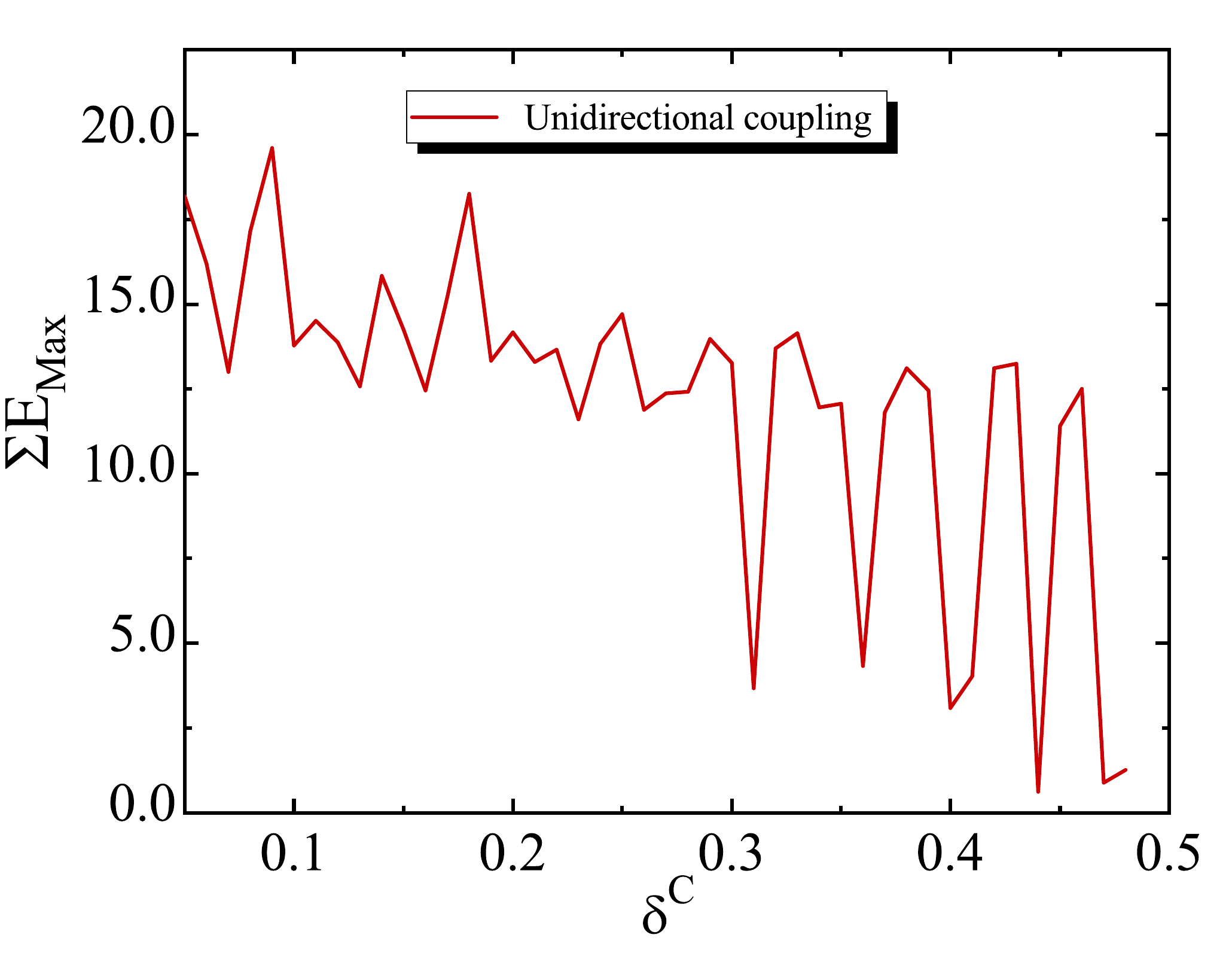}
		\caption{The trend of the maximum system error with the coupling strength for unidirectional coupling. The total error is always oscillating and does not converge to zero. Therefore unidirectional coupling does not synchronize the three Chua circuits.}
		\label{DXEMAX}
	\end{figure}

	For $\delta^C=0.3$, it is completely impossible to synchronize the three circuits (see Fig. \ref{fig:11} (A) , (B) , (C)); For $\delta^C=0.48$, it is only possible to achieve a state of approximate synchronization between some two circuits in part of the time period, but there is still a certain error (see Fig. \ref{fig:12} (A) , (B) , (C)). And this approximate synchronization does not last long, after which it becomes unsynchronized again. Because of the limitation of the definition of capacitor coupling strength, $\delta^C<\delta_{max}^C=0.5$. Therefore, we can assume that the unidirectional coupling of the capacitor cannot fully synchronize the three Chua circuits.Similarly, we have calculated the sum of the maximum system errors $\sum E_{Max}$ for unidirectional coupling (Fig. \ref{DXEMAX}). the results show that when coupling the Chua circuits unidirectionally through capacitors, they cannot be perfectly synchronized, no matter how large the value of the coupling capacitance is taken.

	When using the capacitor closed-loop coupling of the three Chua circuits, we choose the same parameters $\delta^C=0.3$ and $\delta^C=0.48$. The synchronization case is shown in Fig. \ref{fig:11} (D) , (E) , (F) and Fig. \ref{fig:12} (D) , (E) , (F) . It can be seen that both coupling strengths can fully synchronize the three Chua circuits. In fact, we have calculated in Sec. \ref{sec:3} that for a system of Chua circuits coupled by closed-loop capacitors, the entire circuit system can be fully synchronized when the coupling strength satisfies $\delta^C>\delta_{critical}^C$. The critical value of the coupling strength d depends on the basic circuit parameters of the selected Chua circuit.
	
	Thus, the  advantage of closed-loop coupling is, If the available circuit elements are limited to capacitors, the unidirectional coupling does not allow more than two Chua circuits to be fully synchronized. With closed-loop coupling, the entire circuit system can be fully synchronized with a smaller capacitance, and the overall efficiency is much higher than with one-way coupling.

	This is because the closed-loop coupling is essentially an additional negative feedback loop compared to the unidirectional coupling. The Chua circuit itself has nonlinear characteristics, and the negative feedback loop can reduce the degree of nonlinearity of the system by adjusting the voltage or current to make it more stable. Without the negative feedback loop, the system consisting of three Chua circuits would be an unstable system, with the frequency and amplitude of the output signal varying over time. However, when a negative feedback loop is added, the frequency and amplitude of the output signal can be controlled by adjusting the feedback coefficient to stabilize it around a specific value. Specifically, the negative feedback loop compares the output signal of the chaotic circuit with the input signal, then amplifies the difference and inputs it back into the chaotic circuit. This feedback action suppresses the instability of the chaotic circuit and stabilizes its output signal within a specific range.
	
	\section{Circuit Simulation Experiment}
	\label{sec3}
	
	In the previous discussion, we calculated the numerical results of synchronous control of three Chua circuits using capacitive closed-loop coupling. In this section, we will further verify the synchronization between the Chua circuits through simulation experiments.

	We use the circuit simulation software to build the simulation diagrams of the three Chua circuits via capacitive  closed-loop coupling  (see Appendix.\ref{app:1}).The circuit component parameters are chosen as Eq. (\ref{equ:7}). The results of the circuit simulation are shown in  Fig.\ref{fig:sy2} It is confirmed that the Chua circuit is chaotic and can excite the double vortex attractor at this parameter.
	
\begin{figure}[ht]
		\centering
		\includegraphics[width=80mm]{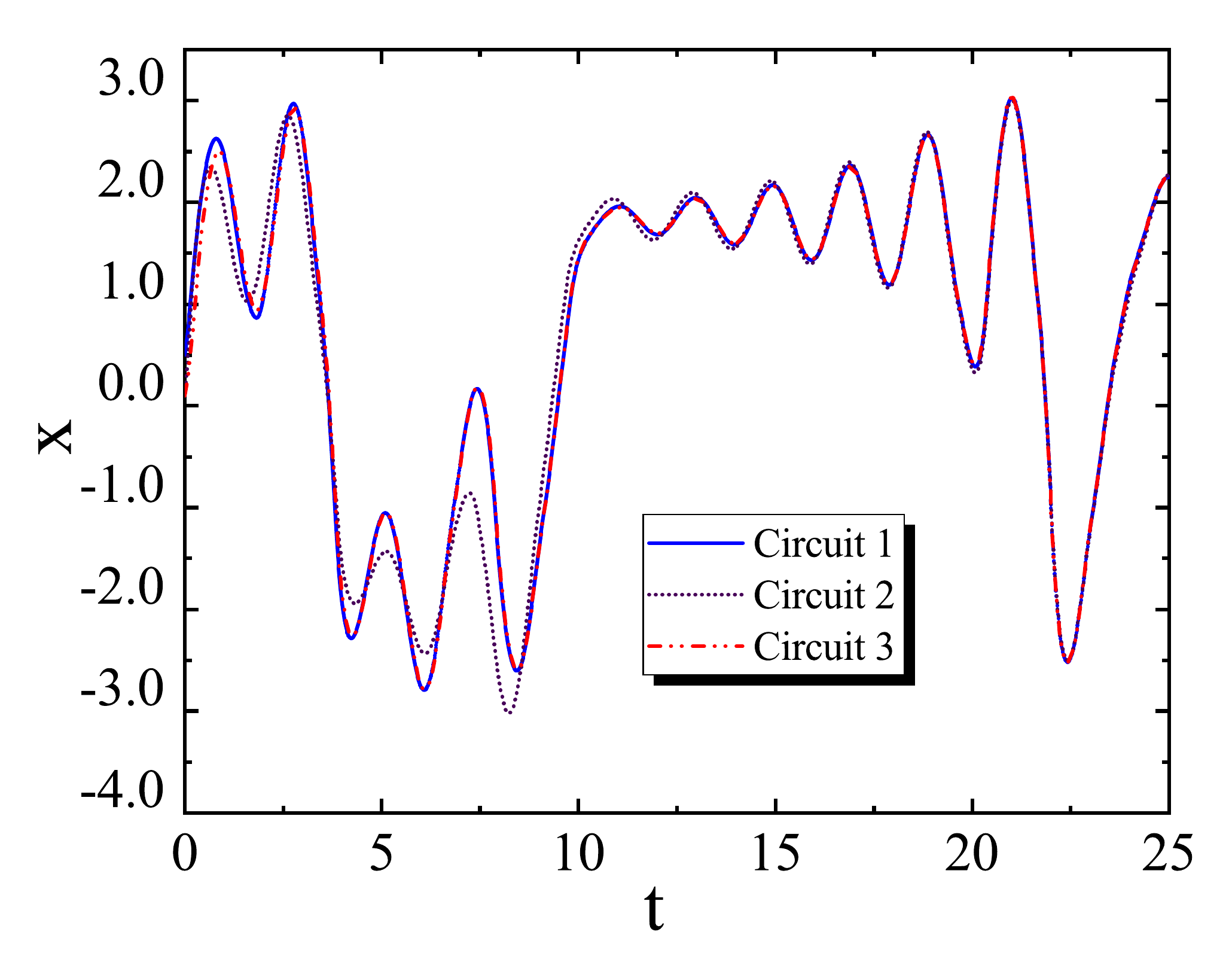}
		\caption{The time series evolution of the output of the "$x$" side of the Chua circuit is plotted. The results show that the time series of the three circuits do not coincide at the beginning due to the different initial values we set for the circuits. However, as time goes on, the outputs of the "$x$" side of the three circuits gradually coincide, which also means that the three Chua circuits are synchronized.}
		\label{fig:16}
	\end{figure}

	Afterwards, we observe the synchronization of the three circuits. When coupling by capacitance, the coupling capacitance $C_c=3.33$nF is chosen and the time series of the $x$ outputs of the three circuits are observed using a four-channel oscilloscope (see Fig. \ref{fig:sy1}). It can be seen that the three curves gradually overlap, which also means that the three Chua circuits gradually become fully synchronized. We can also numerically calculate the time series at the $x$ output. The coupling capacitance $C_c=3.33$ nF corresponds to the coupling strength $\delta^C=0.2$, and the numerical results are shown in Fig. \ref{fig:16} The numerical results are in general agreement with the experimental results.

	Supported by the experimental results, we further determine the feasibility and advantages of making it possible to achieve full synchronization through closed-loop coupled Chua circuits.
	
	\section{Summary}
	\label{sec4}
	In this paper, we have studied the synchronization process of three Chua circuits when they are closed-loop coupled through capacitance. Three Chua circuits with the same circuit parameters and different initial conditions, which are in a chaotic state and can excite a double vortex attractor, are considered. Starting from the circuit equations of the closed-loop coupling of ctapacitance, a dimensionless system equation is constructed, which is the key to the subsequent numerical calculations.
	
	After that, we calculate the feasibility of fully synchronizing the three circuits when capacitance is coupled, and the advantages of closed-loop coupling. And, we determined the coupling strength threshold $\delta_{critical}^C$ for capacitive closed-loop coupling by calculating the maximum total error $\sum E_{Max} $ of the system in a certain time series versus the coupling strength.When the coupling strength $\delta^C \in [\delta_{critical}^C , \delta_{max}^C]$, the capacitive closed-loop coupling can synchronize the three Chua circuits completely, while $\delta^C \in [0 ,\delta_{critical}^C]$ cannot synchronize the circuits. It is also found that the capacitive one-way coupling does not synchronize the circuit completely, regardless of the value of the coupling strength. In particular, if one of the three coupling parameters $\delta^C \in [0 ,\delta_{critical}^C]$, the other $\delta^C \in [\delta_{critical}^C , \delta_{max}^C]$ with the appropriate value can also synchronize the circuit completely. This is the advantage of capacitive closed-loop coupling over other coupling methods.

	Finally, we observed the time series at the $x$ outputs of the three circuits through circuit simulation experiments and compared them with the numerical results again. The feasibility and correctness of the closed-loop coupled Chua circuit using capacitance is further verified. This also provides a new coupling method for the synchronous control of nonlinear circuits.

	\begin{acknowledgments}
		This work is supported by the National Natural Science Foundation of China under Grants No. 12065014 and No. 12047501, and by the Natural Science Foundation of Gansu province under Grant No. 22JR5RA266. We acknowledge the West Light Foundation of The Chinese Academy of Sciences, Grant No. 21JR7RA201.
	\end{acknowledgments}

	\appendix
	\section{Circuit Simulation Diagram}
	\label{app:1}
	\begin{figure*}[h]
		\centering
		\includegraphics[width=0.7\linewidth]{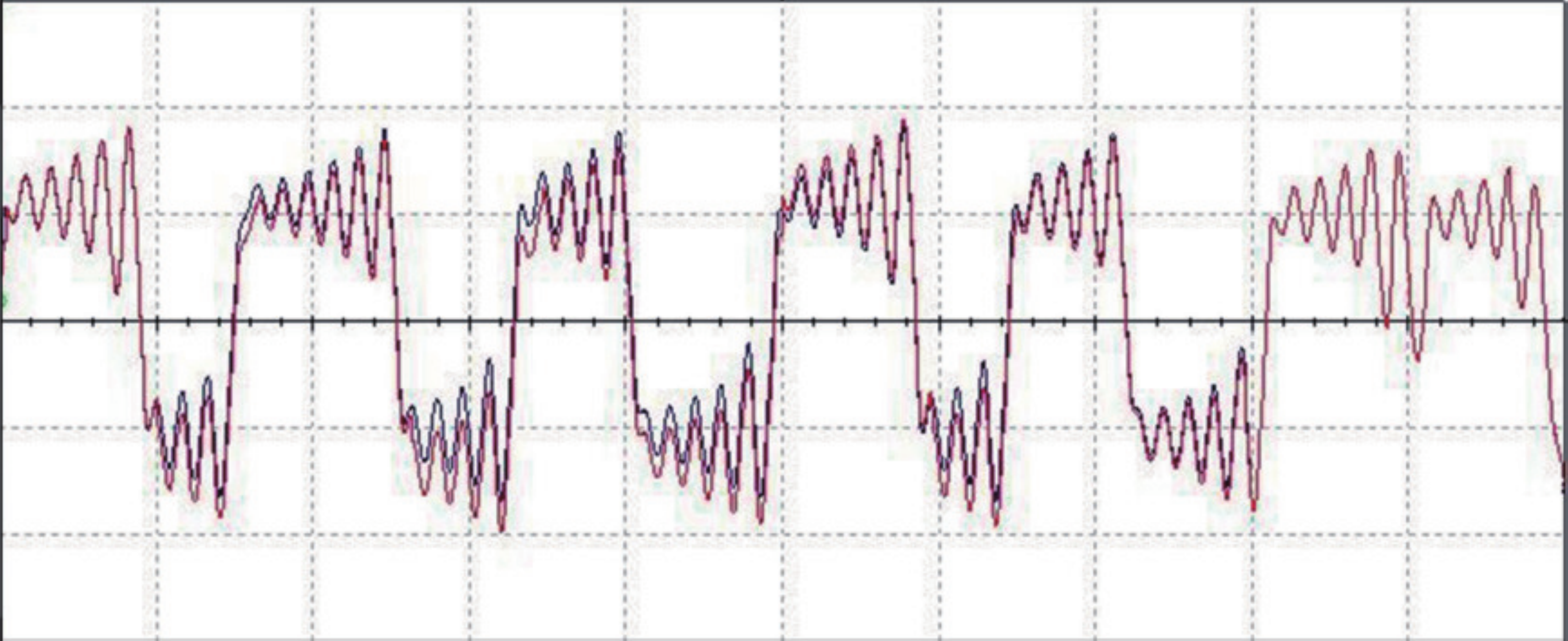}
		\caption{Voltage output time series in simulation}
		\label{fig:sy1}
	\end{figure*}
	\begin{figure*}[h]
		\centering
		\includegraphics[width=1.0\linewidth]{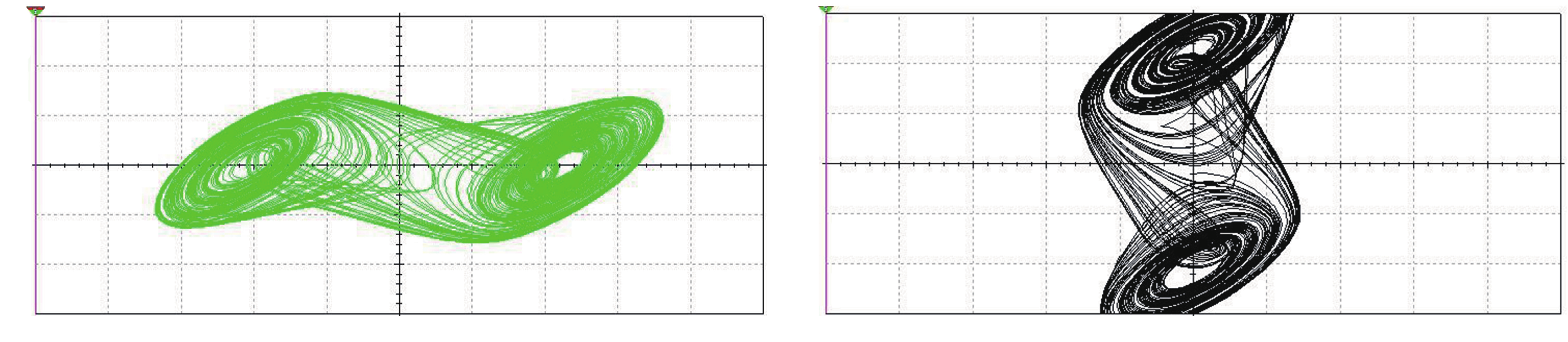}
		\caption{Attractor evolution diagram in simulation}
		\label{fig:sy2}
	\end{figure*}
	\begin{figure*}[h]
		\centering
		\includegraphics[width=1.0\linewidth]{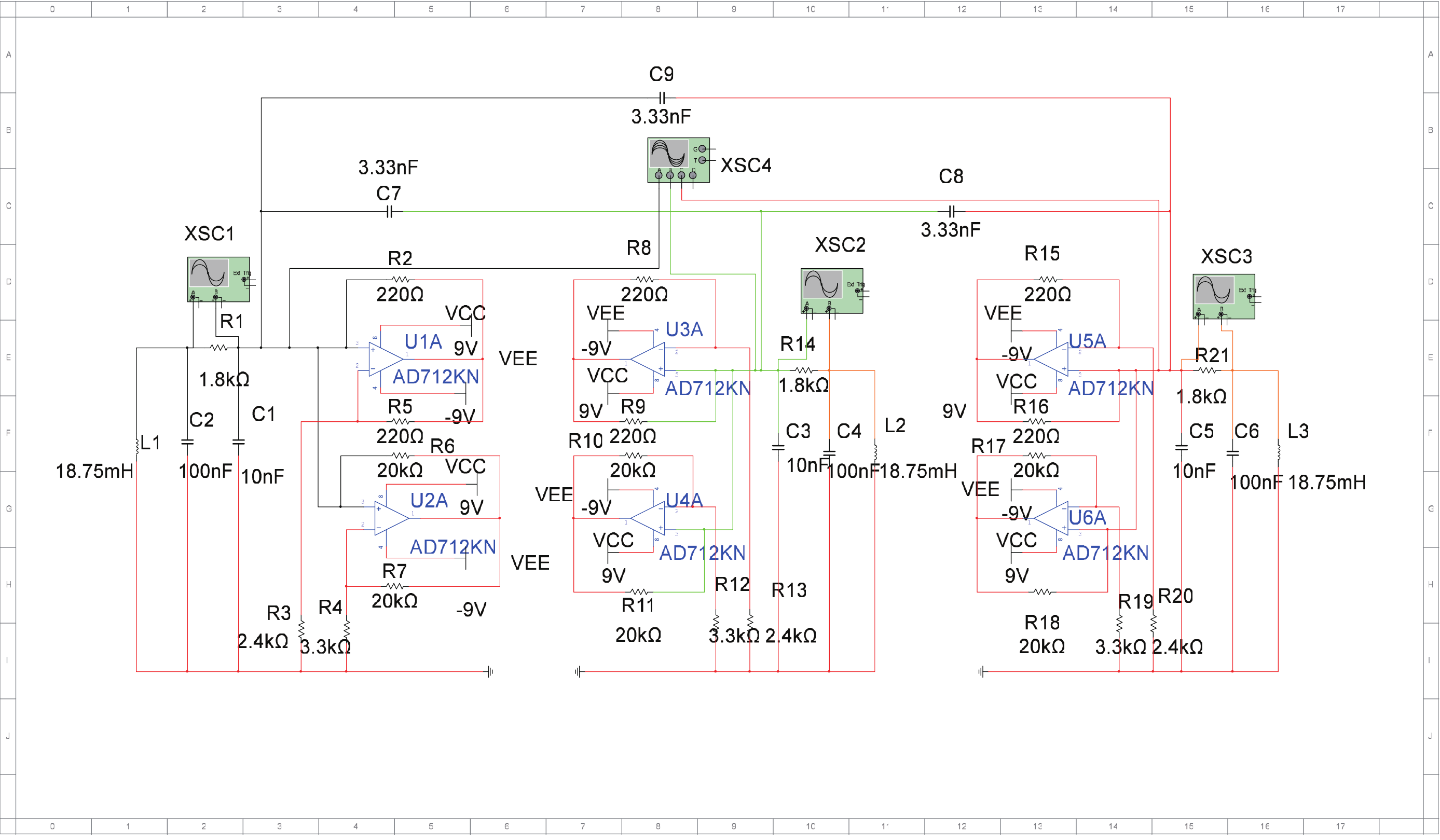}
		\caption{Capacitive closed-loop coupling circuit simulation diagram}
		\label{fig:drfz}
	\end{figure*}
	
\end{document}